\pageno=1                                      
\newcount\notenumber

\def\fnote{\advance\notenumber by 1 \footnote{$^{\the\notenumber}$}}
 
%
%
%
\font\ninerm=cmr9
\font\eightrm=cmr8
\font\sixrm=cmr6
\font\ninei=cmmi9
\font\eighti=cmmi8
\font\sixi=cmmi6
\skewchar\ninei='177 \skewchar\eighti='177 \skewchar\sixi='177
\font\ninesy=cmsy9
\font\eightsy=cmsy8
\font\sixsy=cmsy6
\skewchar\ninesy='60 \skewchar\eightsy='60 \skewchar\sixsy='60

\font\ninebf=cmbx9
\font\eightbf=cmbx8
\font\sixbf=cmbx6
\font\ninett=cmtt9
\font\eighttt=cmtt8
\hyphenchar\tentt=-1 
\hyphenchar\ninett=-1
\hyphenchar\eighttt=-1
\font\ninesl=cmsl9
\font\eightsl=cmsl8
\font\nineit=cmti9
\font\eightit=cmti8
\newskip\ttglue
\def\tenpoint{\def\rm{\fam0\tenrm}%
  \textfont0=\tenrm \scriptfont0=\sevenrm \scriptscriptfont0=\fiverm
  \textfont1=\teni \scriptfont1=\seveni \scriptscriptfont1=\fivei
  \textfont2=\tensy \scriptfont2=\sevensy \scriptscriptfont2=\fivesy
  \textfont3=\tenex \scriptfont3=\tenex \scriptscriptfont3=\tenex
  \def\it{\fam\itfam\tenit}%
  \textfont\itfam=\tenit
  \def\sl{\fam\slfam\tensl}%
  \textfont\slfam=\tensl
  \def\bf{\fam\bffam\tenbf}%
  \textfont\bffam=\tenbf \scriptfont\bffam=\sevenbf
   \scriptscriptfont\bffam=\fivebf
  \def\tt{\fam\ttfam\tentt}%
  \textfont\ttfam=\tentt
  \tt \ttglue=.5em plus.25em minus.15em
  \normalbaselineskip=12pt
  \let\sc=\eightrm
  \let\big=\tenbig
  \setbox\strutbox=\hbox{\vrule height8.5pt depth3.5pt width0pt}%
  \normalbaselines\rm}
\def\ninepoint{\def\rm{\fam0\ninerm}%
  \textfont0=\ninerm \scriptfont0=\sixrm \scriptscriptfont0=\fiverm
  \textfont1=\ninei \scriptfont1=\sixi \scriptscriptfont1=\fivei
  \textfont2=\ninesy \scriptfont2=\sixsy \scriptscriptfont2=\fivesy
  \textfont3=\tenex \scriptfont3=\tenex \scriptscriptfont3=\tenex
  \def\it{\fam\itfam\nineit}%
  \textfont\itfam=\nineit
  \def\sl{\fam\slfam\ninesl}%
  \textfont\slfam=\ninesl
  \def\bf{\fam\bffam\ninebf}%
  \textfont\bffam=\ninebf \scriptfont\bffam=\sixbf
   \scriptscriptfont\bffam=\fivebf
  \def\tt{\fam\ttfam\ninett}%
  \textfont\ttfam=\ninett
  \tt \ttglue=.5em plus.25em minus.15em
  \normalbaselineskip=10pt 
  \let\sc=\sevenrm
  \let\big=\ninebig
  \setbox\strutbox=\hbox{\vrule height8pt depth3pt width0pt}%
  \normalbaselines\rm}
\def\eightpoint{\def\rm{\fam0\eightrm}%
  \textfont0=\eightrm \scriptfont0=\sixrm \scriptscriptfont0=\fiverm
  \textfont1=\eighti \scriptfont1=\sixi \scriptscriptfont1=\fivei
  \textfont2=\eightsy \scriptfont2=\sixsy \scriptscriptfont2=\fivesy
  \textfont3=\tenex \scriptfont3=\tenex \scriptscriptfont3=\tenex
  \def\it{\fam\itfam\eightit}%
  \textfont\itfam=\eightit
  \def\sl{\fam\slfam\eightsl}%
  \textfont\slfam=\eightsl
  \def\bf{\fam\bffam\eightbf}%
  \textfont\bffam=\eightbf \scriptfont\bffam=\sixbf
   \scriptscriptfont\bffam=\fivebf
  \def\tt{\fam\ttfam\eighttt}%
  \textfont\ttfam=\eighttt
  \tt \ttglue=.5em plus.25em minus.15em
  \normalbaselineskip=9pt
  \let\sc=\sixrm
  \let\big=\eightbig
  \setbox\strutbox=\hbox{\vrule height7pt depth2pt width0pt}%
  \normalbaselines\rm}
%
\def\headtype{\ninepoint}                 
\def\abstracttype{\ninepoint}             
\def\captiontype{\ninepoint}              
\def\footnotetype{\ninepoint}             
\def\refit{\it}                           
\font\chaptitle=cmr10 at 11pt             
\rm                                       

%
%
\parindent=0.25in                         
\parskip=0pt                              
\baselineskip=12pt                        
\hsize=4.25truein                         
\vsize=7.445truein                        
\hoffset=1in                              
\voffset=-0.5in                           

\newskip\sectionskipamount                
\newskip\aftermainskipamount              
\newskip\subsecskipamount                 
\newskip\firstpageskipamount              
\newskip\capskipamount                    
\newskip\ackskipamount                    
\sectionskipamount=0.2in plus 0.09in
\aftermainskipamount=6pt plus 6pt         
\subsecskipamount=0.1in plus 0.04in
\firstpageskipamount=3pc
\capskipamount=0.1in
\ackskipamount=0.15in
\def\sectionskip{\vskip\sectionskipamount}
\def\aftermainskip{\vskip\aftermainskipamount}
\def\subsecskip{\vskip\subsecskipamount} 
\def\firstpageskip{\vskip\firstpageskipamount}
\def\capskip{\hskip\capskipamount}

%
%
\nopagenumbers                            
\newcount\firstpageno                     
\firstpageno=\pageno                      
\newcount\chapno                          

\def\rightheadline{\headtype\phantom{\folio}\hfil\runningtitletext\hfil\folio}
\def\leftheadline{\headtype\folio\hfil\runningauthortext\hfil\phantom{\folio}}
\headline={\ifnum\pageno=\firstpageno\hfil
           \else
              \ifdim\ht\topins=\vsize           
                 \ifdim\dp\topins=1sp \hfil     
                 \else
                     \ifodd\pageno\rightheadline\else\leftheadline\fi
                 \fi
              \else
                 \ifodd\pageno\rightheadline\else\leftheadline\fi
              \fi
           \fi}

\def\bottomnumber{\hss\tenrm[\folio]\hss}
\footline={\ifnum\pageno=\firstpageno\bottomnumber\else\hfil\fi}

%
%
%
%
\outer\def\mainsection#1
    {\vskip 0pt plus\smallskipamount\sectionskip
     \message{#1}\vbox{\noindent{\bf#1}}\nobreak\aftermainskip\noindent}
 
\outer\def\subsection#1
    {\vskip 0pt plus\smallskipamount\subsecskip
     \message{#1}\vbox{\noindent{\bf#1}}\nobreak\smallskip\nobreak\noindent}
 
\def\backup{\nobreak\vskip-\baselineskip\nobreak\vskip-\subsecskipamount\nobreak
}

\def\title#1{{\chaptitle\leftline{#1}}}
\def\name#1{\leftline{#1}}
\def\affiliation#1{\leftline{\it #1}}
\def\abstract#1{{\abstracttype \noindent #1 \smallskip\vskip .1in}}
\def\ref{\noindent \parshape2 0truein 4.25truein 0.25truein 4truein}
\def\caption{\noindent \captiontype
             \parshape=2 0truein 4.25truein .125truein 4.125truein}

\def\footnote#1{\edef\fspafac{\spacefactor\the\spacefactor}#1\fspafac
      \insert\footins\bgroup\footnotetype
      \interlinepenalty100 \let\par=\endgraf
        \leftskip=0pt \rightskip=0pt
        \splittopskip=10pt plus 1pt minus 1pt \floatingpenalty=20000
        \textindent{#1}\bgroup\strut\aftergroup\strut\egroup\let\next}
\skip\footins=12pt plus 2pt minus 4pt 
\dimen\footins=30pc 

%
%

\def\@{\spacefactor 1000}

\def\,{\pcomma} 
\def\pcomma{\relax\ifmmode\mskip\thinmuskip\else\thinspace\fi}

\def\oversim#1#2{\lower0.5ex\vbox{\baselineskip=0pt\lineskip=0.2ex
     \ialign{$\mathsurround=0pt #1\hfil##\hfil$\crcr#2\crcr\sim\crcr}}}

\def\runningtitletext{Orion OB1 }
\def\runningauthortext{F.M. Walter et al.}

\null
\firstpageskip

{\baselineskip=14pt
\title{THE LOW MASS STELLAR POPULATION OF THE}
\title{ORION OB1 ASSOCIATION AND IMPLICATIONS}
\title{FOR THE FORMATION OF LOW MASS STARS} } 

\vskip .3truein
\name{Frederick M.\ Walter}
\affiliation{University at Stony Brook}
\vskip .2truein
\name{Juan M.\ Alcal\'a}
\affiliation{Osservatorio Astronomico di Capodimonte}
\vskip .2truein
\name{Ralph Neuh\"auser}
\affiliation{Max--Planck--Institut f\"ur Extraterrestrische Physik}
\vskip .2truein
\name{Michael Sterzik}
\affiliation{European Southern Observatory}
\vskip .1truein
\leftline{and}
\vskip .1truein
\name{Scott J.\ Wolk}
\affiliation{Harvard--Smithsonian Center for Astrophysics}
\vskip .3truein

\abstract{The OB associations, which are fossil star formation regions,
retain the
end-products of the star formation process in an open, unobscured
environment.
Observations of the low mass stars in OB associations provide a far clearer
picture of the results of the star formation process than do observations of
embedded, on-going star formation regions. 
We review the X--ray and optical surveys of the fossil star formation regions
in the Orion OB1 association. 
Low mass pre-main sequence stars abound not only
in the known regions of recent star formation,
but are also distributed over a much larger volume.
The pre-main sequence stars have a narrow age spread, and the mass function
extends down to substellar masses. The clustering of pre-main sequence stars
around $\sigma$~Ori may represent an evolved version of the
Orion Nebula Cluster. We speculate about the effect of the OB environment
on the initial mass function and the formation of planetary systems like the
Solar System.
}


\mainsection{{I}.~~Star Formation in T and OB Associations \hfil \break }
\backup
A long time ago\fnote{about 4.6 Gyr}, in a part of the Galaxy far far away, our
Sun was born. The isotopic abundances in the chondrules and CaIs (chapter
by Goswami and Vanhala, this volume),
in particular the existence of very short-lived parent
radionuclides, provide clues 
about the environment in which our Sun and its planets were born.
Among the plausible sources for these short-lived nuclides are an AGB
star, a supernova, a Wolf-Rayet star, or a combination thereof
(Cameron 1993; but also see Lee et al.\ 1998 and the chapter
by Glassgold et al., this volume).
If the Sun
formed in close proximity to a supernova or a W-R star, then the Sun most
likely formed in an OB association, and not in the quieter confines of a
T~association. 

The T~associations\fnote{Under T~associations
we include all star forming regions
except those dominated by OB stars. This encompasses such regions 
as Taurus-Auriga, Chamaeleon~I, and $\rho$~Oph, which include B and Ae stars.}
are young
unbound groups of low mass stars with ages of a few million years (Myr),
characterized by dust clouds and T~Tauri stars.
Most of what we know about the formation of low
mass stars is based on observations of T~associations, because they are closer
than OB associations and the stars are easier to study. That only a few
T~Tauri stars were known to be associated with the nearby OB~associations
was taken to mean that low mass stars did not form in great numbers in the
OB~associations. However, we now know that only a small percentage of
the low mass
stars form in T~associations: most form in the OB~associations,
which are striking concentrations of short-lived, bright, high mass stars.

The very different environments of OB and T~associations will influence
the formation of the low mass stars and subsequent evolution of their 
protoplanetary disks. To infer the general properties of
low mass pre-main sequence (PMS) stars and their protoplanetary disks
from observations of T~associations alone may lead to biased conclusions.
Our purpose in this review is to draw attention to the low mass star formation
in OB~associations.

We concentrate here on the low mass stars of the Orion OB1 association
(Brown, Walter, and Blaauw 1998 review the high mass stars).
Our focus is on the more exposed parts of the complex, the
fossil star formation regions (SFRs)
of the Ori~OB1a and Ori~OB1b subassociations. 
We will not discuss Ori~OB1c, which surrounds the sword, or Ori~OB1d, the
Orion Nebula Cluster (ONC), a dynamically young, partially embedded,
active SFR associated with the Orion~A cloud (Hillenbrand 1997) .

We begin with a brief review of low mass PMS stars.
In \S~II we
review the techniques used to identify low mass PMS stars. We discuss the
Orion OB1~association in \S~III and IV,
and speculate about the implications for star and
planet formation in \S~V.

\subsection {A.~~The Low Mass PMS Stars in T Associations}

T~Tauri and other stars of its class are characterized by strong Balmer
emission lines, ultraviolet and infrared continuum excesses, and erratic
variability. They are generally found near dark clouds. Joy (1945) first
identified these stars as a distinct class of objects. Ambartsumian (1947)
concluded that these T~Tauri stars were recently formed stars, and introduced
the term ``T~association'' for their groupings. Herbig (1962) arrived at
similar conclusions. Cohen and Kuhi (1979) created a spectral atlas of the
then-known classical T~Tauri stars (cTTS). Herbig and Bell (1988) have produced
the most current catalog of low mass PMS stars. Recent reviews of the cTTS and
their evolution include those by Basri and Bertout (1993) and Stahler and
Walter (1993). 

In the 20 years since the launch of the {\it EINSTEIN} Observatory, X--ray
imaging observations provided a new view of SFRs. Not
only were some cTTS found to be strong soft X--ray sources, but many other
X--ray sources present in these SFRs were identified optically as low mass PMS
stars. Most of these newly-identified PMS stars lacked the strong line emission
and the UV and near-IR continuum excesses of the cTTS,
and were thought to be post-T Tauri stars
(Herbig 1978) -- stars which have lost their circumstellar material and were
evolving towards the zero-age main sequence (ZAMS). Placement on the
Hertszprung-Russell diagram showed that many of these stars had ages of a few
Myr, comparable to those of the cTTS. Walter (1986) called these the naked
T~Tauri stars (nTTS), stars that had become unveiled, but they are more
commonly referred to as the weak T~Tauri stars (wTTS)\fnote{The
definition of the wTTS is that the 
H$\alpha$ equivalent width is less than 10\AA, while the nTTS lack
evidence for circumstellar material. All nTTS are wTTS; the converse is not
always true.}. 

Spurred in part by the realization that X--ray surveys could reveal a more
complete population of low mass PMS stars, much effort went into studying the
nearby T~associations (e.g., Taurus [Walter et al.\ 1988; Neuh\"auser et al.\
1995b]; Chamaeleon [Feigelson et al.\ 1993; Alcal\'a et al.\ 1995]; Corona
Australis [Walter et al.\ 1997]; $\rho$~Oph [Montmerle et al.\ 1983; Casanova
et al.\ 1995]; Lupus [Wichmann et al.\ 1997; Krautter et al.\ 1997]). These
studies have shown, among other things, that 

\item{$\bullet$}
      The low mass population is dominated, at least at ages of
      more than about 1~Myr, by stars which do not have circumstellar disks
      (Walter et al.\ 1988; Strom et al.\ 1989). \par
\item{$\bullet$} 
      Low mass stars can form either in small clumps (cluster mode) or
      individually (distributed mode) (Strom et al.\ 1990). \par
\item{$\bullet$}
      Most stars form in multiple systems.
      The binary fraction in T~associations seems to
      be twice that of the field  (Leinert et al.\ 1993;
      Ghez et al.\ 1993; Simon et al.\ 1993;
      Kohler and Leinert 1998). \par
\item{$\bullet$}
      Low mass stars may be found far from their natal clouds, or may
      not be associated with any recognizable cloud (Walter et al.\ 1988;
      Alcal\'a et al.\ 1995; Neuh\"auser et al.\ 1995c;
      Kastner et al.\ 1997). \par
\item{$\bullet$}
      The distribution of rotational periods is bi-modal, with both slow
      rotators, whose periods are set by disk-breaking, and a rapidly rotating
      population representing stars which have spun up as they contract along
      their Hayashi tracks (Edwards et al.\ 1993; Bouvier, Forestini, and
      Allain 1997). \par
\item{$\bullet$}
      There is a large spread of stellar ages in T associations. \par

\vskip 0.3cm

The large age spread suggests that the star-formation lifetime of a molecular
cloud is related to the sound-crossing time-scale of the cloud (Herbig 1978),
or to the ambi-polar diffusion timescales. Given the typical $1$ to
$2$~km~s$^{-1}$ velocity dispersion within associations (Herbig 1977, Hartmann
et al.\ 1986, Frink et al.\ 1997), one
would expect the older population to have dispersed far from the cloud. The
lack of these older stars gives rise to the post-T~Tauri problem. Palla and
Galli (1997), however, noted that star formation is not a steady process, but
that it runs slowly and less efficiently in the first few million years of the
cloud life-time. Hence, there should be only few post-TTS, but many coeval cTTS
and nTTS. 

\subsection{B.~~The Low Mass PMS Stars in OB Associations}

While the T~associations produce perhaps a few thousand
stars over their 10-30~Myr lifetimes, the OB associations are far more
productive, though for a shorter interval.
OB~associations are loose, easily identifiable concentrations of bright high
mass stars (see Humphries 1978; Blaauw 1964). Ambartsumian (1947)
showed that their typical mass densities of $<$0.1~M$_\odot$~pc$^{-3}$
are unstable to galactic tidal forces, and therefore they
must be young. This conclusion
is supported by the ages derived from the Hertzprung-Russell
diagrams for these associations. 

Blaauw (1991) reviewed the
nearby OB~associations and their relation to local star formation.
The older OB~associations retain a fossil record of 
star formation processes in a giant molecular cloud. The
very process of formation of the massive stars disperses the giant molecular
cloud and thereby disrupts further star formation. 

One can estimate the importance of OB associations for low mass star formation.
Within 500~pc of the Sun lie 3 OB associations (Ori OB1, Sco-Cen-Lup, and
Per~OB2) and a number of T associations (Tau-Aur, Cha, CrA, Lup, TW~Hya).
Assuming a Miller-Scalo mass function in the OB associations, and counting
stars in the T~associations, one can show that over 90\% of the low
mass stars with ages less than about 10~Myr are likely to be members of
OB~associations. 

It had long been thought that star formation was bi-modal, in the sense that
high mass stars did not form in T~associations, and that low mass stars did not
form in great numbers in OB~associations (e.g., Larson 1986, Shu and Lizano
1988). We now know that the latter is not true. H$\alpha$ surveys suggest that
H$\alpha$-emitting stars are found not only in the actively star-forming parts
of giant molecular clouds (e.g., Haro 1953; Duerr et al.\ 1982; 
Strom et al.\ 1990),
but also abound in the fossil OB~associations (e.g.,
Kogure et al.\ 1989; Nakano et al.\ 1995). 


Walter et al.\ (1994) investigated the low mass
population of the Upper~Sco association (de\,Geus, de\,Zeeuw, and Lub 1989),
a 5~Myr old association at a distance of $\sim$140~pc. Starting with
EINSTEIN X--ray observations, they found a low mass PMS
population, whose properties appear
significantly different than those found in T associations.\par
\item{$\bullet$}
   The space density of PMS stars is higher than that in Taurus by about a
      factor of 3.\par
\item{$\bullet$}
   The low mass stars seem to be coeval, at an age of 1-2~Myr.\par
\item{$\bullet$}
   The low mass PMS population is largely devoid of circumstellar material and
   near-IR excesses, even at this age.\par
\item{$\bullet$}
The distribution of rotational periods is not peaked, suggesting that
      the association is observed during an epoch when all the stars are
      spinning up (Adams et al.\ 1998).\par
\item{$\bullet$}
     Between 10 and 0.3~M$_\odot$, the mass function ${d log N}$/${d log M}$
     is consistent with the field star initial mass function
     (Miller and Scalo 1979).\par

\vskip 0.3cm

Sciortino et al.\ (1998) reached similar conclusions about the Upper Sco
Association based on ROSAT observations.
Brandner and K\"ohler (1998) suggested that the binary
fraction of PMS stars in OB~associations is about half that observed in the
T~associations, and is comparable to that of the field. They note that
the binary fraction may be smaller yet near the OB stars.

\vskip 0.3cm

The nearest giant molecular cloud complex and site of active star
formation is in
Orion -- the Orion A and B clouds and the Orion OB1 association. All
stages of the star formation process can be found here, from deeply
embedded protoclusters to fully exposed OB associations.  The
different modes of star formation occurring in these clouds
(clustered, distributed, isolated) allow us to learn more about the
influence of the environment on the star formation process.

We study the Ori~OB1a and OB1b fossil SFRs because in these
regions star formation is complete, all the stars are visible (few embedded
sources remain), and the stars are at their final masses. Yet these fossil
SFRs are sufficiently young (2-10 Myr) that the full
population remains cospatial, and that spatial substructure in the star
formation process may still be detectable.

\mainsection{{I}{I}.~~How to find low mass PMS stars in associations}
\backup

To study the global processes of low mass star formation, 
one must identify and sample all the populations of low mass PMS stars.
The embedded sources, the cTTS, and the nTTS
may represent different populations, with
different spatial and age distributions. All the populations, or their
evolutionary descendents, are present (and none are hidden) in fossil SFRs.
The cTTS tend to be readily identifiable
either photometrically, because of their variability or their prominent near-IR
continuum excesses, or spectroscopically, through
their strong emission lines. However, the vast majority of the
low mass PMS stars are
not so easily discovered. Their coronal X--ray and chromospheric emission
line fluxes are little stronger than those of active ZAMS stars, they have
no continuum
excesses, and they are no more variable than heavily spotted, active late-type
stars. We review methods of searching for the nTTS and wTTS, and identifying
the complete PMS population of an association. 

\subsection{A.~~X--Ray surveys} 

Low mass PMS stars have X--ray luminosities between 10$^{29}$ and
10$^{31}$ erg~s$^{-1}$, reflecting the strong magnetic activity of these stars.
The X--ray surface flux scales with the stellar mass (Walter 1996), which gives
an apparent rotation-activity correlation because stellar rotation rates
increase with increasing mass. There is no clear evidence that either the
X--ray surface flux or the ratio of the X--ray-to-bolometric flux correlates
with stellar rotation\fnote{This is not surprising. Among the Pleiades G stars
(Stauffer et al.\ 1994), a rotation-activity relation is seen only for v
sin~$i<$15~km~s$^{-1}$. Nearly all PMS stars rotate more rapidly.}. The
surface flux of the PMS stars is somewhat lower than the ``saturated'' value
seen in the rapidly rotating dwarfs in the Pleiades (Stauffer et al.\ 1994). 

The high X--ray luminosities, at a characteristic temperature of about 1~keV,
are easily detectable in short pointed X--ray imaging observations, or in
flux-limited surveys like the ROSAT All-Sky Survey (RASS; eg.
Voges et al.\ 1996). Although most cTTS
have been detected as X--ray emitters, they tend to be fairly heavily absorbed,
so their mean observed fluxes in soft X--rays are lower than those
of the nTTS. Neuh\"auser et al.\ (1995a) argue that the cTTS are also
intrinsically less X--ray-luminous than the nTTS. The detection rate for cTTS
in the RASS is only 15\% (Neuh\"auser et al.\ 1995a). The detection efficiency
of unveiled stars is unknown, but deep pointings seem to
recover 70-80\% of the photometrically-identifiable PMS stars (Walter, Wolk,
and Sherry 1998). These X--ray surveys complement optical H$\alpha$ emission
line surveys that mainly yield cTTS. 

The complete sky coverage of the RASS to a flux-limit of
$\approx 1-2 \times 10^{-13}$~erg~cm$^{-2}$~s$^{-1}$
has permitted unbiased analyses of the spatial distribution of
X--ray active stars, including low mass PMS stars. However, the complete
identification and classification of all detected sources 
requires optical spectroscopy of each of the possible counterparts of an
X--ray source. Neuh\"auser et al.\ (1995b) and Sterzik et al.\ (1995)
established a discrimination criterion, based on X--ray spectral appearance and
f$_x$/f$_v$ of known TTS, for selecting PMS candidates. Sterzik et al.\ (1995)
applied it to map the large-scale spatial distribution of young stars in a
700~deg$^2$ field centered on the Orion SFR. The spatial
distribution of the stellar X--ray sources selected in this way gives a
qualitative idea of the general morphology of the SFR, by
tracing the X--ray active young stellar population\fnote{The main source of
contamination of X--ray surveys is by active ZAMS stars that have X--ray colors
and activity levels similar to those of the PMS stars, and that are expected to
dominate among field star X--ray emitters (ages up to about 10$^8$~years;
Brice\~no et al.\ 1997). Such stars are still young, but not necessarily PMS;
their ages must still be determined from observations of the space motions and
parallaxes. A second source of contamination is active binary stars and RS~CVn
systems, for which high levels of X--ray activity are maintained by tidal
coupling of the close binary system.}. 

The $\sim$10$^{-14}$ erg~cm$^{-2}$~s$^{-1}$ limiting fluxes of existing X-ray
observations, together with the 
f$_x$/f${_v}\sim$10$^{-3}$ typically seen in the
PMS stars, means that the X-ray pointings can select optical counterparts
down to about V=15. At the distance and age of Ori~OB1, this corresponds to
$\approx$ 0.5~M$_\odot$. 

\subsection{B.~~Spectroscopy}
The most certain way to identify low mass PMS stars is by
their optical spectra. 


The principle spectroscopic characteristic of low mass PMS stars is the 
6707\AA\ line of Li~I.
This line is generally considered to be an indicator of youth because the
depletion of
Li in the outer layers of convective low mass stars is very rapid.
Lithium is easily
destroyed by convective mixing in the stellar interiors when the temperatures
at the bottom of the convective layer reaches about  2.5$\times$10$^6$~K
(Bodenheimer 1965; D'Antona and Mazzitelli 1994).
The surface Li abundance is indeed anticorrelated with the stellar age in
convective stars (Duncan 1981). Among the G and K stars, there is no
significant Li destruction on the convective track, so lithium can be
used as a criterion
to identify low mass PMS stars of G and K spectral types. Many PMS
stars do exhibit abundances (log[n$_{\rm Li}$]=3)
consistent with undepleted cosmic material (e.g., 
Basri, Martin, and Bertout 1991). 

PMS lithium
burning is expected among the M stars (e.g., Pinsonneault,
Kawaler, and Demarque 1990). Walter et al.\ (1997) observed this
among the lowest mass PMS stars of the CrA~association, but the depletion
was greater than expected by the models. They noted that this
could be accounted for by the fact that the
presence of an active chromosphere/transition region will over-ionize Li
relative to LTE,
giving an apparent underabundance (Houdebine and Doyle 1995).
In the CrA association, the M stars with the largest H$\alpha$~emission
equivalent widths do exhibit
the smallest apparent Li abundances.

The nTT stars exhibit
strong chromospheric emission at H$\alpha$, Ca~II K\&H and
Mg~II~k\&h. The transition region emission lines are at saturated levels.
The saturated chromospheric H$\alpha$ emission surface flux of order
10$^7$~erg~cm$^{-2}$~s$^{-1}$ (averaged over the star) is sufficient to
cause an emission line above the photosphere only
among the late K and M stars.
While such activity appears commonplace among the
PMS stars, it is not unique and cannot be used to identify a star as
PMS. Walter and Barry (1991) discuss the timescales for the decay of the
chromospheric and coronal emissions.

The cTTS show far stronger levels of
chromospheric and transition region emission, including
the hydrogen Balmer lines, the Ca~II K\&H resonance lines,
the FeII $\lambda$ 4924 \AA~ and the
fluorescence lines of FeI $\lambda\lambda$ 4063 and 4132 \AA\ in emission
(Herbig 1962). 
We now know that much of this flux arises, not in a compact solar-like
chromosphere, but in an extended atmosphere and accretion flows (e.g., Hartmann,
Hewett and Calvet 1994; Muzerolle, Hartmann, and Calvet 1998).
Bastian et al.\ (1983) defined a phenomenological working definition for cTTS 
as having an H$\alpha$ emission equivalent width of $>$5\AA, to discriminate
against dMe stars, and in general
a 10\AA\ H$\alpha$ emission equivalent width is used to discriminate between
the cTTS and the nTT or wTT stars.
The emission line spectra of cTTS also often have forbidden
lines of [S~II] and [O~I] that, together with a broad H$\alpha$ emission line
profile and P~Cygni line profiles,
are diagnostics of strong winds (e.g., Mundt 1984; Hartigan, Edwards, and
Ghandour 1995). 

Radial velocities are
powerful tools for establishing membership in a particular cluster or
association, because the dispersion in radial velocities of an association is
1-2~km~s$^{-1}$ (e.g., Hartmann et al.\ 1986). Proper motions (e.g., Jones and
Herbig 1979; Frink et al.\ 1997)
can also be used to establish likely membership.

\vskip 0.2cm

Star formation regions are big: efficient searches must survey large
solid angles.
Traditionally, objective prism
surveys with photographic plates have been used to survey for H$\alpha$
emission-line objects. Most known cTTS have been discovered in such surveys.
Herbig, Vrba, and Rydgren
(1986) used objective prism techniques at Ca~II~K\&H. They could identify 
emission from M stars, but the bright $\lambda\lambda$3900-4000\AA\
continuua of the K stars greatly reduce
sensitivity for hotter stars. 
Because of the generally low spectral resolution ($>$10\AA ),
objective techniques are not suitable for measuring
relatively weak absorption lines, like the important Li~I line.

Once limited to single objects, high resolution spectroscopy is now commonly
undertaken in survey mode, due to
the maturation of optical fiber technology and multi-object spectroscopy.
One can obtain high resolution spectra of hundreds of objects
within a field as large as a square degree simultaneously.
At these higher resolutions, weak absorption lines such as
Li~I~$\lambda$6707\AA\ can be resolved from the
Ca~I~$\lambda$6717\AA\ line, and radial velocities can be measured for all the
objects in the field. Examples of such data are given by Walter, Wolk,
and Sherry (1998).

\subsection{C.~~Wide field imaging photometry}
In an area with little interstellar and circumstellar
reddening, optical color-magnitude diagrams (CMDs) are a powerful tool for
selecting
PMS stars. Large CCD arrays coupled with small (1-meter class)
telescopes can map out large areas of the sky fairly efficiently.
Star identification algorithms such as DAOPHOT are robust
in rejecting objects with non--stellar point spread functions and
provide very reliable photometric results at the mean stellar densities
observed in the Orion OB1~association (about 2 stars brighter than V=18 per
arcmin$^2$).

In practice, we use the known PMS stars
to define the PMS locus in the V~vs.~V--R or V~vs.~V--I CMD.
As the reddening vector is nearly parallel to the
PMS locus in these CMDs, it is not important to know the reddening in order to
identify the PMS candidates. While the X--ray sources calibrate the PMS locus
to about V=15, the optical photometry extends to V$\sim$20 even in short
exposures.

For example, we have
obtained CCD photometry of a 2200 arcmin$^2$ region in Orion's
belt (Ori OB1b), near $\sigma$~Ori.
Completeness of these images, 
as determined by number counts, was V=22, with some stars 
as faint as magnitude V=23 detected in at least three colors.
The data (Fig.\ 1) bifurcate into two distinct groups: a group of stars,
including most of the X--ray sources, which lies in a diagonal band across the
diagram, and a set
of background stars below the band. These data will be discussed
in \S~IV.


\mainsection{{I}{I}{I}.~~LOW MASS STARS IN ORI OB1: LARGE SCALES}
\backup
The Orion complex covers about 25$^{\circ}$ in declination (including the
$\lambda$~Ori region) and one hour in right ascension.
Surveys for low mass stars over this entire area are important for studies of
the subassociation boundaries, the history of star formation, and the
kinematics of the association and its interactions with the
molecular gas.

\subsection{A.~~H$\alpha$ Surveys }

The Kiso objective prism survey in Orion (Wiramihardja
et al.\ 1989; 1991; 1993; Kogure et al.\ 1989) revealed 1157 H$\alpha$
emission line objects over an area of 150~deg$^2$ from 5.15$^h$ to 5.85$^h$
in right ascension and from -13.0$^{\circ}$ to +2.8$^{\circ}$ in declination, to
a limiting magnitude of V=17.5. The magnitude distribution,
peaked around V=15, supports a classical T~Tauri nature for these objects.
Kogure et al.\
(1992) followed-up with low-dispersion spectroscopic observations
of 34 emission line stars in Ori~OB1b, and concluded that they were
indeed T\,Tauri stars based on H-Balmer and CaII~K~emission lines.
Nakano and McGregor (1995) obtained near-IR photometry for a number of
these stars, and also concluded that they were mostly T\,Tauri stars.


 The spatial distribution of the Kiso H$\alpha$ emission-line objects is shown
 in the left panel of Figure~2. 
 Concentrations of emission-line objects
 coincide with the general location of the NGC~2023 and 
 NGC~2024 clusters, the OB1c, OB1b, and OB1a associations.
 Note that emission line objects are also found far outside
 the limits of the OB1 associations.

\subsection{B.~~ROSAT Survey Results}

The surface density distribution of X--ray-selected PMS star candidates
in a 700~deg$^2$ field around the Orion molecular clouds (Sterzik et al.\ 1995)
shows peaks 
associated with subgroup associations (OB1a, OB1b, OB1c, 
and $\lambda$~Ori).
The spatial extent of the density peaks is consistent 
with dispersal times between 2 and 10~Myrs, the
ages of the stellar components in these regions. 
Surprisingly, 
the largest fraction of the PMS star candidates in this
sample is not in this `clustered'
population but is distributed widely over an area 
many times greater than that of the molecular gas or the OB stars.
A large number of sources are seemingly unrelated to any
molecular clouds or fossil SFR. 
This distributed population has also been found
around other nearby SFR's such as Taurus-Auriga (Wichmann et al.\ 1996),
Lupus (Krautter et al. 1997), and Chameleon (Alcal\'a et al.\ 1995).


To explore the nature of these PMS candidates,
Sterzik et al.\ (1997) extended their earlier analysis
to a larger area around Orion. They interpreted
the spatial distribution of PMS candidates (Fig.\ 3) in the framework of
a stellar population model of the galaxy. 
The 6482 RASS sources in the $\sim$ 
5000 deg$^2$ field include a subsample of 
1467 PMS candidates. 
The densest regions (up to 10~stars/deg$^2$) 
coincide with centers of active star 
formation, e.g., in the Orion nebula region, near $\lambda$~Ori, and in 
well-known star forming sites in the Taurus clouds. 
In all these 
cases, the expected sensitivity for selecting young stars is verified. 
In addition, other clusters (e.g., around NGC~1788 [\S~IV.C]) 
which have not 
previously been recognized as prominent SFR's, are also detected  
with this method, and are likely to harbor a high fraction of PMS stars. 

Figure~3 shows that the Orion and Taurus SFRs are projected against an 
$\sim$10$^\circ$
wide strip of apparently young stars (density $<$1 star/deg$^2$ above
the RASS detection limit).
Although Orion and Taurus are at different distances from the 
Sun, they seem to be connected by a broad lane 
that extends further southeastward. This 
contiguous structure is not symmetric about the galactic plane, but 
rather follows the mean location of the Gould Belt as defined 
by early type stars in this direction (Blaauw 1991). The surface density of 
young star candidates drops down to a background 
value of about 0.1 candidate stars/deg$^2$ 
near $b_{II}=0^{\circ}$, and below that value at high galactic latitudes.

Based on the morphology and surface density distribution of X--ray 
selected young star candidates, and detailed comparisons  with the
predictions  of  a  galactic X--ray population model (Guillout et al.\ 1998),
Sterzik et al.\ (1997) show that the X--ray population consists of 
a mixture of three distinct populations:  
(1) The {\sl clustered population} comprises the  
dense regions associated with sites of active or recent 
star formation (e.g., OB1a, OB1b, OB1c, $\lambda$~Ori, NGC 1788).
(2) The strip connecting Orion and Taurus, which coincides with
Gould's Belt.
(3) The {\sl background population} having a density $\sim$ 
0.1 stars/deg$^2$ near the galactic plane, 
which is likely dominated by ZAMS stars.

\subsection{C.~~Optical followups to the ROSAT Survey}

The spatial distribution of 671 PMS candidates in the RASS sample
is shown in the right panel of Figure~2.
There is an apparent spatial
coincidence between those regions of high X--ray source density and those of
high densities of H$\alpha$ emission objects from the Kiso survey. An
immediate conclusion would be that many of the emission-line objects 
are detected in X--rays, but this is not the case.
Fewer than 5\% of the Kiso emission line stars are coincident with
RASS PMS candidate X--ray sources.
Although the Kiso survey goes deeper than the expected optical
magnitude of PMS stars detectable in the RASS,
over 90\% of the RASS X--ray sources are not coincident with an
emission-line star.
We note that only a handful of emission-line objects
in the $\lambda$~Ori region (Duerr et al.\ 1982) coincide with RASS sources.

The lack of coincidences between the RASS sources and the 
H$\alpha$ emitters
is a consequence of the fact that cTTS are more difficult to detect in
soft X--rays than nTTS, perhaps 
because X--rays are efficiently absorbed in the
dense circumstellar envelopes of cTTS (Walter and Kuhi 1981)
or because nTTS rotate faster than cTTS
(Bouvier et al.\ 1993; 1995). 
The important implication is that most
of the PMS stars are not emission-line objects.

 Alcal\'a et al.\ (1996) observed a spatially unbiased sample of 181 RASS 
 sources, using
 intermediate resolution long slit spectroscopy and photoelectric photometry.
 They identified 112 stars which
 showed Li absorption and a late-type
spectrum. These low mass PMS star candidates have a
spatial distribution indistinguishable from that shown 
 in the right panel of Figure~2.

 More recently, Alcal\'a et al.\ (1998) placed a representative subsample
 of these stars in the HR diagram assuming a distance of 460~pc
 and found that they fall well above the main sequence with typical T~Tauri
 masses and ages ($0.8<M_{star}/M_{\odot}<3.4$; 
 0.2 $< \tau_{age}$(Myr) \ $< 7$).
They found that the stars with
stronger Li~I ($\lambda$ 6707\AA) line tend to concentrate toward
the Orion molecular clouds, but they did not find any other
correlations between the spatial location, age, or other stellar parameter.
The lack of stars with masses less than 0.8$M_{\odot}$ in this sample
is simply due to the flux limit of the RASS, and the
approximately constant f$_x$/f$_v$ of PMS stars. 

 To verify the Li measurements
 (which are blended at moderate resolution), and to obtain radial velocities,
 which can be used to distinguish association members from non-members, 
 Alcal\'a et al.\ (1999) undertook high resolution (R=25,000)
 spectroscopic observations of many of the 112 stars in the Alcal\'a et al.\
(1996) sample.


The distribution of radial velocities of the RASS stars in Orion
   which have strong Li and appear to be PMS is shown in Figure~4.
   The radial velocity distribution is broad with a velocity dispersion of
   about 9 km s$^{-1}$ and apparently shows a double peak, one at
   $\approx$18~km s$^{-1}$ and the other at $\approx$25~km s$^{-1}$.
   The former coincides with the mean radial velocity for Orion, while the
   latter appears to coincide with the radial velocity of the Taurus clouds.
   This suggests that the RASS PMS sample in Orion is a juxtaposition
   of two distinct groups of stars, one associated with the Orion SFR.
   The relation of the other group to the Taurus SFR is unclear, as these stars
   would lie a minimum of 50-80~pc from the currently-active star formation
   in the Taurus clouds. This group might be related to the
   Gould's Belt population (\S~III.B).

 There is no statistical difference in the lithium strength between
 the two radial velocity groups: most of the stars have lithium 
 abundances comparable to those of low mass PMS stars. We conclude that most
 of these stars in the RASS sample are indeed PMS, and that at least some are
 demonstrably members of the Orion association.

\mainsection{{I}{V}.~~LOW MASS STARS IN ORI OB1: SMALL SCALES}
\backup
On scales of a few degrees or less, the Orion OB1 association breaks up into
individual subassociations. Orion is sufficiently young that the
subassociations retain
their individual identities. One can investigate the timescales of the star
formation process, and study the initial mass function and mass segregation.
Aside from the obvious differences in age and total mass, just how similar
are these subassociations?

\subsection{A.~~The Belt}

A series of ROSAT PSPC pointings (Walter 1994) in the belt of Orion revealed
hundreds of X--ray sources, of which many are now confirmed
low mass PMS stars, based on
spectroscopic and photometric followups. We have concentrated our efforts
on the region surrounding $\sigma$\,Ori (Walter, Wolk, and Sherry 1998).
The ROSAT PSPC and HRI observations reveal over 100 X--ray point
sources within 1$^{\circ}$ of $\sigma$\,Ori, a member of Ori~OB1b
and a Trapezium-like system.

Walter, Wolk, and Sherry (1998)
obtained spectra of most of the optical counterparts of
the X--ray sources near $\sigma$~Ori, and of a
randomly-selected sample of nearby stars.
Among these $\sim$300 stars, they identified 104 likely PMS
stars within 30 arcmin of $\sigma$\,Ori. Primary
identification was made on the basis of a strong Li\,I
$\lambda$6707\AA\ absorption line. The H$\alpha$ strengths range from an
emission equivalent width of 77\AA\ in a K1 star to normal
photospheric absorption. The distribution
of radial velocities is strongly peaked at the 25~km s$^{-1}$ velocity
of the OB association (Fig.\ 5).


The color-magnitude diagram (Fig.\ 1) for 0.6\,deg$^2$ surrounding
$\sigma$\,Ori
shows a clear PMS locus, well separated from
the background galactic stars. The narrowness
of the PMS locus suggests coevality, but an age less than the 1.7\,Myr age
of the OB association. 

The spectroscopic sample, which is spatially uniform and statistically complete
to about V=15, has a surface density of 120 PMS stars/deg$^2$ (10$<$V$<$15),
and shows evidence for clustering. The centroid of the
PMS star distribution is centered on the position of $\sigma$\,Ori.
Summation of the stars into radial bins centered on $\sigma$\,Ori
shows that the distribution is flat for the
non-PMS stars, but that the radial distribution of the PMS stars is
peaked at $\sigma$\,Ori. The inferred cluster radius (where the density
of PMS stars reaches zero) is about 0.5\,deg (3.3\,pc).
If all the PMS stars are distributed in this way, then the
space density, in the magnitude
range 12$< V <$19, is about 400 stars per deg$^2$ (or more, if
many stars are multiple).
The total inferred mass of this group of stars is about
half that of the ONC\@. The $\sigma$\,Ori cluster is
the {\it second youngest\/} cluster now known after the ONC, and
may be an evolved analog of the ONC\@.

Wolk (1996) showed that about 30\% of these stars are slow
rotators and 70\% are rapid rotators. Similarly, about 30\% of the
spectroscopically-identified PMS stars have strong H$\alpha$ emission, and
appear to be cTTS.
This suggests that, at an age of about
2~Myr, 30\% of the low mass stars near $\sigma$~Ori
retain their circumstellar disks. This is a higher
fraction than seen in Upper Sco (Walter et al.\ 1994) at a similar age.

The existence of the $\sigma$\,Ori cluster implies that there is substructure
in Ori~OB1b. As
there is no evidence for evolutionary differences between the early-type
stars in Ori~OB1b (Brown et al.\ 1994),
it may be that Ori~OB1b formed through the
merging of several Trapezium-like clusters that formed at more or less
the same time. Observations are underway to test this hypothesis.

\subsection{B.~~Ori OB1a}

Observations of five
other regions in Ori~OB1 reveal similar concentrations of PMS stars, with
space densities (V$<$15)
from about 40 to 150 stars/deg$^2$, comparable to that
near $\sigma$\,Ori. We find the highest space density of PMS stars
within the Ori~OB1a association, near 5$^h$~24$^m$ +1$^{\circ}$.
This sample of stars includes no slow rotators (rotation periods longer than
4 days; Wolk 1996), and only one
classical T\,Tauri star, suggesting that essentially all circumstellar disks
have dissipated by the $\sim$10~Myr age of this association.
The CMD is similar to that for
the $\sigma$~Ori region, except that the 
PMS stars appear older.
For a 330 pc distance (Brown, Walter, and Blaauw 1998), the PMS stars lie
above the 10~Myr locus expected from the age of the B stars. In general, 
the low mass stars in OB associations appear younger than the high mass stars.
This effect is evident in $\sigma$~Ori, in the $\lambda$~Ori region
(Dolan 1998), and in the Upper Sco association (Walter et al.\ 1994).

\subsection{C.~~NGC~1788}

A significant density enhancement of RASS PMS candidates is present
near the reflection nebula NGC~1788, which coincides with the
CO-clump \#13 of Maddalena et al.\ (1986) at  
$\alpha \approx$~76$^\circ$ and $\delta \approx$~-3.5$^\circ$.
This reflection nebula is centered on a cluster of stars and 
illuminated by the B9V~star HD~293815 (Witt and Schild 1986). 
The cluster seems heavily obscured by foreground material, consistent
with gas column densities $> 5 \times 10^{22}$~N$_{H_2}$ 
derived from $^{12}$CO/$^{13}$CO line ratio observations (Knapp et al.\ 1977).
This high column density is likely to hide most X--ray sources
in the cloud itself. 
One cTTS, LkH$\alpha$~333, is known in its vicinity.
A surface density of $\approx$~4~sources/deg$^2$ in
an elongated structure of 3$^{\circ}$ length and 1.5$^{\circ}$ width is
detected in the RASS.
A preliminary analysis of the recent HRI pointing
resulted in the discovery of more than 50 additional X--ray sources, probably
embedded in the molecular cloud.
The relatively small width of the density enhancement would indicate
a diffusion age of less than 5~Myr, if these stars have formed in a central
cluster with a dispersion velocity of 2~km s$^{-1}$.
More support for the existence of a young cluster
near NGC~1788 follows from recent near-infrared imaging by Dougados et al.\
(in preparation).

Unlike the fossil SFRs, NGC~1788 appears to be a region
of on-going star formation far from the Orion A and B clouds. Its relation to
the Orion complex is not known.

\subsection{D.~~Run-Away T Tauri Stars and the Distributed Population} 

Not all PMS stars are found in or near molecular clouds:
RASS observations have shown that a distributed population of PMS stars
appears to be a common characteristic of SFRs.
While such a distributed population may have formed locally
(e.g., Feigelson 1996), Sterzik et al.\ (1995) 
suggested that the PMS stars in the distributed population were ejected
from their birth clouds with high velocities. They called such stars
``run-away TTS'' (raTTS).
Few-body encounters can happen early in the lifetime of a multiple protostellar
system, so that they are also of relevance in establishing the fraction
of binary (and triple) PMS stars.
Many on-cloud PMS stars are multiples, so
the multiplicity must be established very early in the
PMS phase (e.g., Leinert et al.\ 1993, Ghez et al.\ 1993, Mathieu 1994 and
chapter by Mathieu et al., this volume).
In star-star or star-cloud encounters such as those studied by
Sterzik and Durisen (1995), ejected raTTS
are either single stars or close binaries, and should be on average
less massive than average TTS.
Gorti and Bhatt (1996) modeled the ejection
of protostars in encounters of protostars with clouds, and found
that some protostars can be ejected in such a way.
Kroupa (1995) showed that several percent of the members of a cluster
as rich as the Trapezium can be ejected by close encounters with
velocities exceeding 5 km s$^{-1}$.

The characterization of any individual star as an raTTS requires 
detailed knowledge of their space motions and ages.
There are a few well-studied stars in Orion
whose space motions, locations, and ages indicate that they may well
be raTTS. These include: Par 1540 (Marschall and Mathieu 1988) which may
have been ejected from the ONC $\sim 10^{5}$ years ago;  
Par 1724 (V1321 Ori; Neuh\"auser et al.\ 1998), which
is moving north at about
10 km s$^{-1}$, and may also have been ejected
from the ONC $\approx 10^{5}$ yrs ago; and RXJ0511.2+1031
(Magazz\`u et al. 1997; Neuh\"auser et al.\ 1997)
which may have originated in the $\lambda$~Ori region.

These examples show that there are indeed very young stars far away
from star forming clouds, whose space motions point back to those clouds.
RaTTS certainly do exist, and if they are numerous the RaTTS may be the
bridge between the star formation, occurring on small scales, and the observed 
large scale distribution of PMS stars observed after a few million years.

\mainsection{{V}.~~APPLICATIONS AND IMPLICATIONS}
\backup
\subsection{A.~~The Low End of the Mass Function }
Very low mass (VLM) objects -- the lowest mass stars as well as substellar mass
objects -- are quite bright when young. 
A 2~Myr old star at the hydrogen burning limit (0.08~M$_\odot$) is only
about 10 times less luminous than the Sun (Burrows et al.\ 1997).
At the distance of Orion, such stars will have V$\approx$19, which is easily
observable with small telescopes. By the age
of the Pleiades, VLM stars have faded by about 3 magnitudes
at V, and are some 200-400K cooler. So while IR studies and large
telescopes (e.g., Stauffer, Hamilton, and Probst 1994; Zapatero Osorio,
Rebolo, and Martin 1996) are required to detect VLM objects in
clusters like $\alpha$ Persei and the Pleiades, a more modest approach will
yield similar results in nearby SFRs.

At the 2~Myr age of Ori~OB1, a star near the H-burning limit will have
V--R$\sim$1.2 (Baraffe et al.\ 1998). 
Our deep photometry of the $\sigma$~Ori field
(Fig.\ 1) shows that the empirical PMS
locus seems to continue 
into the brown dwarf regime. To determine whether the red colors are
intrinsic, or are due to extreme reddening,
Wolk (in preparation) has obtained low resolution spectra of several of
these objects. The filled circle in Figure~1 redward of the
theoretical brown dwarf limit 
is of spectral type M6 with minimal reddening. M6 is
the spectral type of 
the Pleiades brown dwarfs PPl15 and Teide~1, confirmed by the 
``Lithium test'' (Basri, Marcy, and Graham 1996). Since a younger object
of the same spectral type must be less massive, it is highly likely that our
object is of substellar mass. 
It is unlikely that all the other very red objects
are subject to extreme reddening.
The 2200 arcmin$^2$ area included in Figure~1 contains 8 additional
brown dwarf candidates both fainter and redder than the candidate
discussed above.

Color-magnitude diagrams of 2200 arcmin$^2$ around $\delta$~Ori and of
1100 arcmin$^2$ around $\epsilon$~Ori reveal 9 photometric brown dwarf
candidates. This is a density of about 12 brown dwarfs per
deg$^2$, and suggests that there may be an abundance of substellar mass objects
in OB associations.

\subsection{B.~~The Initial Mass Function}
The easiest time to determine the initial mass function (IMF), the relative
numbers of stars as a function of mass,
is early in the life of an association or a cluster, before high
mass stars burn out, and before dynamical
friction segregates the masses and ejects the lower mass systems. 
Studies of unobscured fossil SFRs -- the OB associations --
afford all of the advantages of studying the IMF in embedded clusters
(chapter by Meyer et al., this volume), and none of the disadvantages
encountered
in highly obscured regions where the stars may not yet have reached their
final masses. 
At ages of a few Myr, one can readily see to --
and below -- the hydrogen-burning limit. The ability to identify PMS stars
independent of their activity levels gives the opportunity
to identify a (statistically) complete population of PMS stars. In
OB associations we can directly observe and measure the
IMF from O stars through substellar mass objects.

The published data suggest that there is a universal slope to the mass function
in associations, and that it is similar to the field star mass function
(Miller and Scalo 1979). Walter and Boyd (1981) found that the mass function
in Taurus approximated the field star mass function from a few solar masses
to about 0.3~M$_\odot$; Walter et al.\ (1994) found the same in Upper Sco.
Hillenbrand (1997) found the mass function in the ONC to agree overall with the
Miller and Scalo IMF, though with some differences in detail. This 
universality should
not be surprising, because most stars are born in associations, which disperse
to populate the field. However, the question of whether there are small
differences in the mass function between associations remains open.

One place where differences may exist is at the low mass end of the mass
function. Brown dwarfs
appear to be rare in T~associations (e.g., Stauffer et al.\ 1991), although
recently Luhman et al.\ (1997) and Neuh\"auser and Comer\'on (1998) reported
finding brown dwarfs in the $\rho$~Oph and Cha~I SFRs,
respectively. But there appears to be a high density of VLM objects in Ori OB1b.
This may be a simple consequence of the differences between the
T~and OB
association environments. A low mass protostar in a T~association is able to
accrete mass for a time set by the accretion process 
and the local environment.
However, in an OB association, the mass accretion can be terminated 
abruptly as the local cloud is disrupted by 
nearby massive stars (see Walter et al.\ 1994), either due to winds or
supernovae. So while in a T~association a
protostar may accumulate a significant fraction of the mass within its
original Jean's radius, a low mass
protostar in an OB association may never accrete that
last fraction of a solar mass, and VLM stars may end up with substellar masses.
OB associations may be the place to search for brown dwarfs. 

There is some
evidence that this may indeed be the case. In the Taurus T association, the
typical PMS star is spectral type K7-M0, suggesting a mass function peaking near
0.3-0.4~M$_\odot$. Walter et al.\ (1994) found the typical low mass PMS star in
Upper Sco to be early M, and found that the mass function extended to
0.2~M$_\odot$. Hillenbrand (1997) finds a mass function peaking at 0.2~M$_\odot$
in the ONC. 
If the mass functions are different in OB and T associations, it will
have profound consequences for where one looks for brown dwarfs, or perhaps for
large planets.

\subsection{C.~~Disk survival times and implications for planets}

We cannot yet directly detect planets orbiting PMS stars in Orion, or in
any SFR. We cannot yet detect terrestrial-size planets anywhere.
We can detect circumstellar disks, which are likely to be a necessary
ingredient for planet formation, but we do not know how the disk turns into
planets. Models which qualitatively explain the distribution of planets
in our solar system fail to predict the existence of giant planets close
to their stars (but see chapter by Lin et al., this volume).
An observer of stars and disks can only speculate about how
planets might fare in an OB environment, but can describe those conditions
under which planet formation must proceed.

Elsewhere in this volume, Hollenbach et al.\ discuss the
evaporating disks seen in
HST images of the Orion Nebula. In a simple picture, 
longer-lived disks provide more time for planets to form, and
disks in T~associations may survive longer than disks in OB associations.
One needs about a million years
to form a Jupiter (see review by Wuchterl
et al.\ in this volume). In the ONC, with
a nominal age of about 0.5~Myr, about half the low mass PMS stars seem to have
disks (Hillenbrand 1997). In the older $\sigma$~Ori cluster (nominal age
1.7~Myr), about 30\% seem to retain disks, based on using rotation periods
and H$\alpha$ emission as disk proxies, while by the age of Ori~OB1a there is
little evidence for disk survival. The disk survival times
do appear to be short in OB associations.

Short disk survival times may not impede the formation of terrestrial
planets: small bodies can form very quickly, and need not accumulate
in the presence of gas (Weidenschilling and Cuzzi 1993; Lissauer and Stewart
1993). Indeed, short disk survival times may be advantageous
for terrestrial-sized planets. Lin, Bodenheimer, and Richardson (1996)
suggest that giant planets tend to migrate inwards in the presence of a disk,
sweeping ahead of them all planetesimals which may have formed
in the inner planetary system (see
also the reviews in this volume by Lin et al. and by Ward and Hahn).
Short disk survival times may prevent such
orbital migration, and protect small inner planets. Thus one could make a case
that planetary systems like our own may be most likely to form in
environments like OB associations.

\subsection{D.~~Summary}

There is a wealth of knowledge to be gained about the global processes of star
formation from studying the low mass population of OB associations. 
The IMF is best determined in fossil star formation regions, because all
the stars are readily countable, and have reached their final masses. The
ages of the low mass stars can be estimated more accurately (subject to
systematic uncertainties in the evolutionary models) than can the ages of 
massive stars already on the main sequence, permitting studies of the
coevality of star formation across the association.
The apparent difference in ages
between the high- and low-mass stars may provide information about the triggers
and timescales of low mass star formation.
The radial velocities of the
low mass stars can often be measured more precisely than can those of the
rapidly rotating O and B stars, and can provide more definitive measures of the
kinematics of OB associations. The spatial distribution of the numerous
low mass stars can yield insights into the substructuring of the associations.

Low mass stars, revealed by X--ray and H$\alpha$ surveys, abound in the fossil
SFRs of Orion OB1. 
The PMS stars not only concentrate
in the known subassociations, but are also distributed over a much larger
volume than are the OB stars. Optical photometry shows that the PMS
stars have a narrow age spread, and extend down to substellar masses. 
The high density of brown dwarf candidates in Ori OB1b may be a consequence
of star formation in the OB environment. The ONC
is not the only young cluster in Orion; $\sigma$~Ori has its own cluster.
Subclustering may be common: the belt of Orion may be the amalgamation
of ONC-like systems which formed at about the same time. 

Most low mass
stars in our galaxy likely formed in large OB associations like the Orion OB1
association. There is circumstantial evidence that our Sun may have formed in
an OB association 4.6~Gyr ago. If so, then star formation in the OB 
environment does not preclude the formation of planetary systems, and we can be
optimistic that planetary systems like our own are common in the galaxy.

\vfill\eject
\null

\vskip .5in
\centerline{\bf FIGURE CAPTIONS}
\vskip .25in

\caption{Figure 1.\capskip  
The color--magnitude diagram for all stars (dots) in the 2200 square arcminute
region centered on $\sigma$ Ori.
Effective temperatures are derived from Bessell (1995).
The dashed lines trace the expected pre-main sequence locus.
The vertical dotted line indicates the brown dwarf cut--off derived in Baraffe
et al.\ (1998) with corrections for ``missing opacity'' and mean reddening.
The ``X's'' indicate X--ray sources confirmed to be PMS by their spectra.
Filled
circles are photometrically--selected and spectroscopically--confirmed PMS
candidates. ``T's'' indicate classical T Tauri stars among these confirmed
candidates.  Open circles are photometric candidate PMS stars whose V--R and
R--I colors are consistent
with PMS nature (uncircled dots within the PMS locus have discrepant R--I
colors), but for which we have no spectra. The A$_V$=1 reddening vector
is indicated. 
The clear separation of the PMS stars from the general distribution of
background stars: demonstrates the efficacy of using wide-field optical
photometry to identify low mass PMS stars in OB~associations.
}


\caption{Figure 2.\capskip Spatial distribution of the H$\alpha$ 
 emission line objects found in the Kiso survey (left panel) and 
 of the ROSAT all-sky survey X--ray sources in Orion (right panel). 
 The position of the OB1 associations as well as the NGC~2023 and 
 NGC~2024 clusters (Zinnecker et al.\ 1993)
 are indicated in the left panel. The outlines 
 of the CO survey by Maddalena et al.\ (1986) are also show. The 
 dashed square in each panel indicates the extents of the Kiso survey.
While the strongest density enhancements in both the H$\alpha$ and X--ray
source densities are associated with the Orion~A cloud, significant
numbers of sources extend out beyond the molecular cloud.
}

\caption{Figure 3.\capskip 
Large-scale spatial distribution of candidate young stars in Orion 
and Taurus-Auriga. The X--ray sources (dots) have been selected from the ROSAT 
All-Sky Survey. Their space density is color-coded, in units of 
X--ray sources per square degree.
Contours of molecular clouds (Maddalena et al.\ 1986) are overlayed.
The dashed lines show the galactic plane and galactic longitude=180$^\circ$.
The regions of 
enhanced young star candidate density 
south of the galactic plane coincide with Gould's Belt.

The Orion Nebula Cluster is the strongest enhancement at the center of the
image. The enhancements immediately to the north of the ONC are Ori~OB1b and
Ori OB1a. NGC~1788 is west of Ori~OB1b.
The Pleiades cluster is visible at the northwestern edge of the field. 
}

\caption{Figure 4.\capskip Radial velocity distribution of 
 the RASS PMS candidates observed with high-resolution. The radial 
 velocity of the Orion association is about 25~km~s$^{-1}$; the peak near
 18~km~s$^{-1}$ is consistent with the radial velocity of the nearby
 Taurus SFR (Hartmann et al.\ 1986). This suggests that the RASS survey
has detected two discrete populations of low mass PMS stars.
}

\caption{Figure 5.\capskip The radial velocity
  distribution of stars within 30 arcmin of $\sigma$~Ori.
  We determined radial velocities by cross-correlating
  the spectra with those of the dusk or dawn sky.
  Uncertainties are about $\pm$5~km~s$^{-1}$ for spectra with high S/N.
  The spectroscopically-identified PMS stars (solid histogram) are
  well-fit as a Gaussian distribution of mean 25~km~s$^{-1}$ with
  $\sigma$=5~km~s$^{-1}$. The secondary peak at 12~km~s$^{-1}$ is due
  to a systematic shift of M star velocities, and may be an
  artifact of using a sky spectrum as a velocity template. The radial
  velocities of the other stars in the sample (dotted histogram) have
  a mean of 31~km~s$^{-1}$ with $\sigma$=37~km~s$^{-1}$. 
  It is clear from the radial velocities that the PMS stars in this field
are all members of the same association.
}

\vfill\eject


\null

\vskip .5in
\centerline{\bf REFERENCES}
\vskip .25in

\ref{Adams, N.R. Walter, F.M., Wolk, S.J. 1998. Rotation Periods of Low-Mass
   Stars of the Upper Scorpius OB Association.
   {\refit Astron.\ J.\/} 116: 237--244.

\ref{Alcal\'a J.M., Chavarr\'ia-K C., Terranegra L. 1998. 
   On the nature of the ROSAT X--ray selected weak-line T Tauri stars in Orion.
   {\refit Astron.\ Astrophys.\/} 330: 1017--1028.}

\ref{Alcal\'a J.M., Covino E., Torres G., Sterzik M.F., Pfeiffer M.,
    Neuh\"auser R. 1999.
   High-resolution spectroscopy of ROSAT weak line T Tauri stars in Orion.
   {\refit Astron.\ Astrophys.\/} submitted.}

\ref{Alcal\'a J.M., Krautter J., Schmitt J.H.M.M., Covino E., Wichmann R.,
   Mundt R. 1995. A study of the Chamaeleon star forming region from the
   ROSAT all-sky survey:I. X--ray observations and optical identifications
   {\refit Astron.\ Astrophys.\ Suppl.\/} 114: 109--134.}

\ref{Alcal\'a J.M. , Terranegra L., Wichmann R., Chavarria C., Krautter J.,
   Schmitt J.H.M.M., Moreno-C M.A., de Lara E., Wagnmer R.M. 1996.
   New weak-line T Tauri stars in Orion from the ROSAT all-sky survey
   {\refit Astron.\ Astrophys.\ Suppl.\/} 119: 7--24, (A96).}

\ref{Ambartsumian V.A., 1947. {\refit Stellar Evolution and Astrophysics}\/,
    Acad. Sci. Armen., Yerevan.}


\ref{Baraffe, I., Chabrier, G., Allard, F., Hauschildt, P.H. 1998.
   Evolutionary models for solar metallicity low-mass stars:
   mass-magnitude relationships and color-magnitude diagrams.
   {\refit Astron.\ Astrophys.\/} 337: 403--412.}

\ref{Basri, G. and Bertout, C. 1993. T Tauri stars and their accretion disks.
   In {\refit Protostars \& Planets {I}{I}{I}\/}, eds.\ E. H. Levy and
   J. I. Lunine (Tucson: Univ.\ of Arizona Press), pp.\ 543--566.}

\ref{Basri, G. Marcy, G.W., Graham, J. R. 1996. Lithium in Brown Dwarf
   Candidates: The Mass and Age of the Faintest Pleiades Stars.
   {\refit Astrophys.\ J.\/} 458: 600--609.}

\ref{Basri G., Martin E., Bertout C. 1991. 
   The lithium resonance line in T Tauri stars. 
   {\refit Astron.\ Astrophys.\/} 252: 625--638.}

\ref{Bastian U., Finkenzeller U., Jascheck C., Jascheck M.. 1983.
   The definition of T Tauri and Herbig Ae/Be stars.
   {\refit Astron.\ Astrophys.\/} 126: 438--439.}

\ref{Bessell, M. 1995. The Faint End of the Stellar Luminosity Function.
   {\refit Pub.\ Astron.\ Soc.\ Aust.\/} 12: 128-- .}    

\ref{Blaauw, A. 1964. The O associations in the solar neighborhood.
   {\refit Ann.\ Rev.\ Astron.\ Astrophys.\/} 2: 213--247.}

\ref{Blaauw, A. 1991. OB associations and the fossil record of star formation.
   in {\refit The Physics of Star Formation and Early Stellar Evolution}, eds.\
   C.J. Lada and N.D. Kylafis (Dordrecht: Kluwer), pp.\ 125--154.}

\ref{Bodenheimer P. 1965. Studies in stellar evolution. II. Lithium depletion
   during the pre-main-sequence contraction.
   {\refit Astrophys.\ J.\/} 142: 451--461.}

\ref{Bouvier J. Cabrit S., Fern\'andez M., Mart\'in E.L., Matthews J.M. 1993.
   COYOTES I. Multisite UBVRI photometry of 24 pre-main sequence 
   stars of the Taurus-Auriga cloud.
   {\refit Astron.\ Astrophys.\ Suppl.\/} 101: 485--498.}

\ref{Bouvier J., Covino E., Kovo O., Mart\'in E.L., Matthews J.M.,
   Terranegra L., Beck S.C. 1995.
   COYOTES II: spot properties and the origin of photometric period variations
   in T Tauri stars.
    {\refit Astron.\ Astrophys.\/} 229: 89--107.}

\ref{Bouvier, J., Forestini, M., and Allain, S. 1997. The angular momentum
    evolution of low-mass stars.
    {\refit Astron.\ Astrophys.\/} 326: 1023--1043.}

\ref{Brandner, W. and K\"ohler, R. 1998. Star Formation Environments and the
   Distribution of Binary Separations.
    {\refit Astrophys.\ J.\ Lett.\/} 499: L79--L82.}


\ref{Brice\~no C., Hartmann L.W., Stauffer J.R., Gagn\'e M., Stern R.A., 
    Caillault J.-P. 1997.  X--Rays Surveys and the Post-T Tauri Problem.
    {\refit Astron.\ J.\/} 113: 740--751.}

\ref{Brown, A.G.A., de Geus, E.J., de Zeeuw, P.T. 1994. The Orion OB1
    association. 1: Stellar content.
    {\refit Astron.\ Astrophys.\/} 289: 101--120.}

\ref{Brown, A.G.A.B., Walter, F.M., Blaauw, A. 1998. The Large-Scale
    Distribution and Motions of Older Stars in Orion. In 
    {\refit The Orion Complex Revisited\/}, eds. A. Burkert and M. McCaughrean,
    in press.}

\ref{Burrows, A., Marley, M., Hubbard, W.B., Lunine, J.I., Guillot, T.,
    Saumon, D., Freedman, R., Sudarsky, D., Sharp, C. 1997. 
    A Nongray Theory of Extrasolar Giant Planets and Brown Dwarfs.
    {\refit Astrophys.\ J.\/} 491: 856--875.}    

\ref{Cameron, A.G.W. 1993. Nucleosynthesis and Star Formation.
    In {\refit Protostars \& Planets {I}{I}{I}\/}, eds.\ E.H. Levy and
    J.I. Lunine (Tucson: Univ.\ of Arizona Press), pp.\ 47--73.}

\ref{Casanova, S., Montmerle, T., Feigelson, E.D., Andr\'e, P. 1995.
    ROSAT X--ray sources embedded in the rho Ophiuchi cloud core.
    {\refit Astrophys.\ J.\/} 439: 752--770.}

\ref{Cohen, M. and Kuhi, L.V. 1979. Observational studies of pre-main
    sequence evolution. 
    {\refit Astrophys.\ J.\ Suppl.\/}  41: 743--843.}

\ref{D'Antona F. and Mazzitelli I. 1994. New pre-main-sequence tracks for
   M$\le2.5M_\odot$ as tests of opacities and convection model. 
   {\refit Astrophys.\ J.\ Suppl.\/} 90: 467--500.}

\ref{Dolan, C. 1998. poster presented at Protostars and Planets {I}{V} meeting.}


\ref{Duerr R. Imhoff C.L., Lada C.J. 1982. Star formation in the $\lambda$
   Orionis Region. I. The distribution of young objects.
   {\refit Astrophys.\ J.\/} 261: 135--150.}

\ref{Duncan D.K. 1981. 
    Lithium abundances, K line emission and ages of nearby solar type stars.
    {\refit Astrophys.\ J.\/} 248: 651--669.}

\ref{Edwards, S., Strom, S.E., Hartigan, P., Strom, K.M., Hillenbrand, L.A.,
    Herbst, W., Attridge, J., Merrill, K.M., Probst, R., Gatley, I. 1993.
    Angular momentum regulation in low-mass young stars surrounded by
    accretion disks.
    {\refit Astron.\ J.\/} 106: 372--382.}

\ref{Feigelson E.D. 1996. 
   Dispersed T Tauri stars and galactic star formation
   {\refit Astrophys.\ J.\/} 468: 306--322.}

\ref{Feigelson, E.D., Casanova, S., Montmerle, T., Guibert, J. 1993.
   ROSAT X--Ray Study of the Chamaeleon I Dark Cloud. I. The Stellar Population.
   {\refit Astrophys.\ J.\/} 416: 623--646.}

\ref{Frink S., R\"oser S., Neuh\"auser R., Sterzik M.F. 1997.
    New proper motions of pre-main sequence stars in Taurus-Auriga.
    {\refit Astron.\ Astrophys.\/} 325: 613--622.}

\ref{de Geus, E.J., de Zeeuw, P.T., Lub, J. 1989. Physical parameters of
    stars in the Scorpio-Centaurus OB association.
    {\refit Astron.\ Astrophys.\/} 216: 44--61.}

\ref{Ghez A.M., Neugebauer G., Matthews K. 1993.
   The multiplicity of T Tauri stars in the star forming regions 
   Taurus-Auriga and Ophiuchus-Scorpius: A 2.2 micron speckle imaging survey.
   {\refit Astron.\ J.\/} 106: 2005--2023.}

\ref{Gorti, U., Bhatt, H.C. 1996.
   Dynamics of embedded protostar clusters in clouds.
   {\refit Mon.\ Not.\ Roy.\ Astron.\ Soc.\/} 278: 611--616.}

\ref{Guillout, P., Sterzik, M.F., Schmitt, J.H.M.M., Motch, C., Egret, D.,
    Voges, W., Neuh\"auser, R. 1998.
    The large-scale distribution of X--ray active stars.
    {\refit Astron.\ Astrophys.\/} 334: 540--544.}

\ref{Haro G. 1953. H$\alpha$ emission stars and peculiar objects
   in the region of the Orion Nebula.
    {\refit Astrophys.\ J.\/} 117: 73--83.}

\ref{Hartigan P., Edwards, S. Ghandour, L. 1995. 
   Disk Accretion and Mass Loss from Young Stars.
   {\refit Astrophys.\ J.\/} 452: 736--768.}

\ref{Hartmann L., Hewett R., Calvet, N. 1994. Magnetospheric accretion models
   for T Tauri stars. I: Balmer line profiles without rotation.
   {\refit Astrophys.\ J.\/} 426: 669--687.}

\ref{Hartmann L., Hewett R., Stahler S., Mathieu R.D. 1986.
     Rotational and radial velocities of T Tauri stars.
     {\refit Astrophys.\ J.\/} 309: 275--293.}

\ref{Herbig G.H. 1962. 
     The properties and problems of T Tauri stars and related objects.
     {\refit Adv.\ Astron.\ Astrophys.\/} 1: 47--103.}

\ref{Herbig G.H. 1977. Radial velocities and spectral types of T Tauri stars.
    {\refit Astrophys.\ J.\/} 214: 747--758.}

\ref{Herbig G.H. 1978. The post T Tauri stars. in {\refit Problems of
    physics and evolution of the universe\/}, ed. L. Myrzoyan
    (Yervan: Academy of Sciences of the Amenian S.S.R.) pp.\ 171--183.}

\ref{Herbig, G.H. and Bell, K.R. 1988. Third Catalog of Emission Line Stars of
    the Orion Population. {\refit Lick Obs. Bull.\/} \# 1111.

\ref{Herbig G.H., Vrba F.J., Rydgren A.E. 1986. 
    A spectroscopic survey of the Taurus-Auriga dark clouds for
    pre-main-sequence stars having Ca~II~H,K emission.
    {\refit Astron.\ J.\/} 91: 575--582.}

\ref{Hillenbrand, L.A. 1997. On the Stellar Population and Star-Forming
    History of the Orion Nebula Cluster.
    {\refit Astron.\ J.\/} 113: 1733--1768.}

\ref{Houdebine, E.R. and Doyle, J.G. 1995. Observation and modelling of
    main sequence star chromospheres. IV. The chromospheric contribution
    to LiI lines in active dwarfs.
    {\refit Astron.\ Astrophys.\/} 302: 861--869.}

\ref{Humphries, R.M., 1978. Studies of Luminous Stars in Nearby galaxies. I. 
    Supergiants and O Stars in the Milky Way.
    {\refit Astrophys.\ J.\ Suppl.\/}  38: 309--350.}

\ref{Jones B.F. and Herbig G.H. 1979. Proper motions of T Tauri variables and
    other stars associated with the Taurus-Auriga dark clouds.
    {\refit Astron.\ J.\/} 84: 1872--1889.}

\ref{Joy, A.H. 1945. T Tauri variables.
    {\refit Astrophys.\ J.\/} 102: 168--195.}

\ref{Kastner J.H., Zuckerman, B., Weintraub, D.A., Forveille, T. 1997. X--ray 
    and molecular emission from the nearest region of recent star formation.
   {\refit Science} 277: 67--71.}

\ref{Knapp, G.R., Kuiper, T.B.H., Knapp, S.L., Brown, R.L. 1977. 
   CO observations of galactic reflection nebulae.
   {\refit Astrophys.\ J.\/} 214: 78--85.}

\ref{Kohler, R. and Leinert, Ch. 1998.
    Multiplicity of T Tauri stars in Taurus after ROSAT.
    {\refit Astron.\ Astrophys.\/} 331: 977--988.}

\ref{Kogure T., Yoshida S., Wiramihardja S.D., Nakano, M., Iwata, T., 
    Ogura, K. 1989. Survey observations of emission-line stars in the
    Orion region. II. The Kiso area A-0903.
    {\refit Pub.\ Astron.\ Soc.\ of Japan\/} 41: 1195--1213.}

\ref{Kogure, T., Ogura, K., Nakano, M., Yoshida, S. 1992.
    Spectroscopic observations of emission-line stars in the
    Orion. I - Ori OB 1b region.
    {\refit Pub.\ Astron.\ Soc.\ of Japan\/} 44: 91--99.}

\ref{Krautter J., Wichmann R., Schmitt J.H.M.M., Alcal\'a J.M., Neuh\"auser R.,
   Terranegra L. 1997.  New weak-line T Tauri stars in Lupus. 
   {\refit Astron.\ Astrophys.\ Suppl.\/} 123: 329--352.}

\ref{Kroupa P. 1995. Star cluster evolution, dynamical age estimation and the
   kinematical signature of star formation
   {\refit Mon.\ Not.\ Roy.\ Astron.\ Soc.\/} 277: 1522--1540.}

\ref{Larson, R.B. 1986. Bimodal star formation and remnant-dominated galactic 
   models.
   {\refit Mon.\ Not.\ Roy.\ Astron.\ Soc.\/} 218: 409--428.}

\ref{Lee, T., Shu, F.H., Shang, H. Glassgold, A.E., Rehm, K.E. 1998.
   Protostellar Cosmic Rays and Extinct Radioactivities in Meteorites.
   {\refit Astrophys.\ J.\/} 506: 898--912.}

\ref{Leinert C., Zinnecker H., Weitzel N., Christou, J., Ridgway, S.T.,
   Jameson, R., Haas, M., Lenzen, R. 1993.
   A systematic approach for young binaries in Taurus.
   {\refit Astron.\ Astrophys.\/} 278: 129--149.}

\ref{Lin, D.N.C., Bodenheimer, P., Richardson, D.C. 1996. Orbital migration
   of the planetary companion of 51 Pegasi to its present location.
   {\refit Nature} 380: 606--607.}

\ref{Lissauer, J.J. and Stewart, G.R. 1993. Growth of Planets from
     Planetesimals.
    In {\refit Protostars \& Planets {I}{I}{I}\/}, eds.\ E.H. Levy and
    J.I. Lunine (Tucson: Univ.\ of Arizona Press), pp.\ 1061--1088.}

\ref{Luhman, K.L., Liebert, J., Rieke, G.H. 1997. Spectroscopy of a Young
   Brown Dwarf in the rho Ophiuchi Cluster 1. 
   {\refit Astrophys.\ J.\ Lett.\/} 489: L165--L168.}

\ref{Maddalena R.J., Morris M., Moscowitz J., Thaddeus P. 1986.
   The large system of molecular clouds in Orion and Monoceros.
   {\refit Astrophys.\ J.\/} 303: 375--391.}

\ref{Magazz\`u A., Martin E.L., Sterzik M.F., Neuh\"auser R., Covino E., 
   Alcal\'a J.M. 1997.
   Search for young low-mass stars in a ROSAT selected sample south of the
   Taurus-Auriga molecular clouds.
   {\refit Astron.\ Astrophys.\/} 124: 449--467.}

\ref{Marschall, L.A., and Mathieu, R.D. 1988. Parenago 1540 - A
   pre-main-sequence double-lined spectroscopic binary near the Orion Trapezium.
    {\refit Astron.\ J.\/} 96: 1956--1964.}

\ref{Mathieu R.D. 1994. Pre-Main-Sequence Binary Stars.
   {\refit Ann.\ Rev.\ Astron.\ Astrophys.\/} 32: 465--530.}

\ref{Miller G.E. and Scalo J.M. 1979. The initial mass function and stellar
   birthrate in the solar neighborhood.
   {\refit Astrophys.\ J.\ Supp.\/} 41, 513--547.}

\ref{Montmerle Th., Koch-Miramond L., Falgarone E., Grindlay J. E. 1983. 
    Einstein observations of the rho Ophiuchi dark cloud: an X--ray
    Christmas tree.
    {\refit Astrophys.\ J.\/} 269: 182--201.}

\ref{Mundt, R. 1984. Mass loss in T Tauri stars - Observational studies of
    the cool parts of their stellar winds and expanding shells. 
    {\refit Astrophys.\ J.\/} 280: 749--770.}

\ref{Muzerolle, J. Hartmann, L. Calvet, N. 1998. Emission-Line Diagnostics of
    T Tauri Magnetospheric Accretion. I. Line Profile Observations.
    {\refit Astron.\ J.\/} 116: 455--468.}

\ref{Nakano, M. and McGregor, P.J. 1995. in {\it Future Utilisation of
   Schmidt Telescopes}, eds. J. Chapman, R. Cannon, S. Harrison, and
   B. Hidayat, {\refit ASP Conf.\ Ser.\/} 84: 376--378.}

\ref{Nakano, M., Wiramihardja, S.D., Kogure, T. 1995. Survey Observations of
    Emission-Line Stars in the Orion Region V. The Outer Regions.
    {\refit Pub.\ Astron.\ Soc.\ of Japan\/} 47: 889--896.}

\ref{Neuh\"auser R. and Comer\'on, F. 1998. ROSAT X-ray Detection of a Young
    Brown Dwarf in the Chamaeleon I Dark Cloud.
    {\refit Science\/} 282: 83--85.}

\ref{Neuh\"auser, R., Sterzik, M.F., Schmitt, J.H.M.M., Wichmann, R.,
   Krautter, J. 1995a. ROSAT survey observation of T Tauri stars in Taurus.
   {\refit Astron.\ Astrophys.\/} 297: 391--417.}

\ref{Neuh\"auser, R., Sterzik, M.F., Schmitt, J.H.M.M., Wichmann, R.,
   Krautter, J. 1995b. Discovering new weak-line T Tauri stars in Taurus-Auriga
   with the ROSAT All-Sky Survey.
   {\refit Astron.\ Astrophys.\/} 295: L5--L8.}

\ref{Neuh\"auser, R., Sterzik, M.F., Torres, G., Martin, E.L. 1995c.
   Weak-line T Tauri stars south of Taurus.
   {\refit Astron.\ Astrophys.\/} 299: L13--L16.}

\ref{Neuh\"auser R., Torres G., Sterzik M.F., Randich S. 1997.
   Optical high-resolution spectroscopy of ROSAT detected late-type stars
   south of the Taurus molecular clouds.
   {\refit Astron.\ Astrophys.\/} 325: 647--663.}

\ref{Neuh\"auser R., Wolk S.J., Torres G., Preibisch Th., Stout-Batalha N.M., 
   Hatzes A., Frink S., Wichmann R., Covino E., Alcal\'a J.M., Brandner W., 
   Walter F.M., Sterzik M.F., 1998. Optical and X--ray monitoring, Doppler
   imaging, and space motion of the young star Par 1724 in Orion. 
   {\refit Astron.\ Astrophys.\/} 334: 873--894.}

\ref{Palla F. and Galli F. 1997. Post-T Tauri Stars: A False Problem.
    {\refit Astrophys.\ J.\ Lett.\/} 476: L35--L38.}

\ref{Pinsonneault, M.H., Kawaler, S.D., and Demarque, P. 1990,
    Rotation of low-mass stars - A new probe of stellar evolution.
    {\refit Astrophys.\ J.\ Supp.\/} 74: 501--550.}

\ref{Sciortino, S., Damiani, F., Favata, F., Micela, G. 1998. An X--ray study
   of the PMS population of the Upper Sco-Cen association.
   {\refit Astron.\ Astrophys.\/} 332: 825--841.}


\ref{Shu, F.H., and Lizano, S. 1988. The evolution of molecular clouds. In
    {\refit Interstellar matter, Proceedings of the second Haystack Observatory
    meeting}, eds. J.M. Moran and P.T.P. Ho (New York: Gordoin and Breach)
    pp.\ 65--74.}

\ref{Simon, M., Ghez, A.M., Leinert, Ch. 1993. Multiplicity and the ages of
    the stars in the Taurus star-forming region.
    {\refit Astrophys.\ J.\ Lett.\/} 408: L33--L36.}


\ref{Stauffer, J.R., Caillault, J.-P., Gagne, M., Prosser, C.F., Hartmann, L.W.
    1994. A deep imaging survey of the Pleiades with ROSAT.
    {\refit Astrophys.\ J.\ Suppl.\/} 91: 625--657.}

\ref{Stauffer, J.R., Hamilton, D., Probst, R.G. 1994. A CCD-based search for
    very low mass members of the Pleiades cluster.
    {\refit Astron.\ J.\/} 108: 155--159.}

\ref{Stauffer, J.R., Herter, T., Hamilton, D., Rieke, G.H., Rieke, M.J.,
     Probst, R.G., Forrest, W. 1991. 
     Spectroscopy of Taurus cloud brown dwarf candidates.
     {\refit Astrophys.\ J.\ Lett.\/} 367: L23--L26.}

\ref{Stahler S.W. and Walter, F.M. 1993. Pre-main sequence evolution and the
   population.
   In {\refit Protostars \& Planets {I}{I}{I}\/}, eds.\ E. H. Levy and
   J. I. Lunine (Tucson: Univ.\ of Arizona Press), pp.\ 405--428.}

\ref{Sterzik, M.F., Alcal\'a, J.M., Neuh\"auser, R., Schmitt, J.H.M.M. 1995.
    The spatial distribution of X--ray selected T-Tauri stars. I. Orion.
    {\refit Astron.\ Astrophys.\/} 297: 418--426.}

\ref{Sterzik M.F. and Durisen, R.H. 1995.
    Escape of T Tauri stars from young stellar systems.
    {\refit Astron.\ Astrophys.\/} 304: L9--L12.}

 \ref{Sterzik M.F., Alcal\'a J.M., Neuh\"auser R., Durisen R.H. 1997.
     The Large-Scale Distribution of X--ray Sources in Orion.
     {\refit The Orion Complex Revisited\/}, eds. A. Burkert and M. McCaughrean,
    in press.}

\ref{Strom K.M., Wilkin F.P., Strom S.E., Seaman R.L.  1989.
   Lithium abundances among solar-type pre-main-sequence stars.
   {\refit Astron.\ J.\/} 98: 1444--1450.}

\ref{Strom, K.M., Strom, S.E., Wilkin, F.P., Carrasco, L., Cruz-Gonzalez, I.,
    Recillas, E., Serrano, A., Seaman, R.L., Stauffer, J.R., Dai, D.,
    Sottile, J. 1990. A study of the stellar population in the Lynds 1641 dark
    cloud. IV - The Einstein X--ray sources.
    {\refit  Astrophys.\ J.\/} 362, 168--190.}

\ref{Voges, W., Boller, T., Dennerl, K., Englhauser, J., Gruber, R.,
    Haberl, F., Paul, J., Pietsch, W., Tr\"umper, J., Zimmermann, H.U. 1996. 
    Identification of the rosat all-sky survey sources. In
    {\refit R\"ontgenstrahlung from the Universe}, eds.\ H.U. Zimmermann, J.
    Tr\"umper, and H. Yorke, MPE Report 263, pp.\ 637--640.}


\ref{Walter F.M. 1986. 
    X--ray sources in regions of star formation I. The Naked T Tauri stars.
    {\refit Astrophys.\ J.\/} 306: 573--586.}

\ref{Walter F.M. 1994. Star formation in Orion (the constellation). In
    {\refit The Soft X--Ray Cosmos}, eds. E.M. Schlegel and R. Petre.
    (New York: API Conf.\ Proc.\ 313), pp.\ 282--284.}

\ref{Walter, F.M. 1996. Coronal magnetic activity in low mass pre-main
    sequence stars. In {\refit Magnetodynamic Phenomena in the Solar
    Atmosphere: Prototypes of Stellar Magnetic Activity}, eds.\ Y. Uchida,
    T. Kosugi, and H.S. Hudson. (Dordrecht: Kluwer), pp.\ 395--396.}

\ref{Walter, F.M. and Barry, D.C. 1991. Pre- and main-sequence evolution of
    solar activity. In {\refit The Sun in Time},
    eds.\ C.P. Sonnett, M.S. Giampapa, and M.S. Matthews
    (Tucson: Univ.\ of Arizona Press), pp.\ 633--657.}

\ref{Walter, F.M. and Boyd, W.T. 1981. Star formation in Taurus-Auriga --
    The high-mass stars.
    {\refit Astrophys.\ J.\/} 370: 318--323.}

\ref{Walter F.M., Brown A., Mathieu R.D., Myers P.C., Vrba F.J. 1988.
    X--ray sources in regions of star formation III. Naked T Tauri stars
    associated with the Taurus-Auriga complex.
    {\refit Astron.\ J.\/} 96: 297--325.}

\ref{Walter F.M., Kuhi L. 1981. The smothered coronae of T Tauri stars.
    {\refit Astrophys.\ J.\/} 250: 254--261.}

\ref{Walter, F.M., Wolk, S.J., and Sherry, W. 1998. The $\sigma$~Orionis
    cluster. In {\refit Cool Stars, Stellar Systems, and the Sun X.}, eds.
    R. Donahue and J. Bookbinder (ASP Conf.\ Ser.\/), CD-1793.}

\ref{Walter, F.M., Vrba F.J., Mathieu R.D., Brown A., Myers P.C., 1994.
   X--ray sources in regions of star formation III. The low mass stars of the
   Upper Scorpius Association.
   {\refit Astron.\ J.\/} 107: 692--719.}

\ref{Walter, F.M., Vrba, F.J., Wolk, S.J., Mathieu, R.D., Neuh\"auser, R.
    1997. X--Ray Sources in Regions of Star Formation. VI. The R CrA
    association as viewed by EINSTEIN.
    {\refit Astron.\ J.\/} 114: 1544--1554.}

\ref{Weidenschilling, S.J. and Cuzzi, J.N. 1993. Formation of Planetesimals in
     the Solar Nebula.
    In {\refit Protostars \& Planets {I}{I}{I}\/}, eds.\ E.H. Levy and
    J.I. Lunine (Tucson: Univ.\ of Arizona Press), pp.\ 1031--1060.}

\ref{Wichmann, R., Krautter, J., Covino, E., Alcal\'a, J.M., Neuh\"euser, R.,
    Schmitt, J.H.M.M. 1997. The T Tauri star population in the Lupus star
    forming region.
    {\refit Astron.\ Astrophys.\/}  320: 185--195.}

\ref{Wichmann R., Krautter J., Schmitt J.H.M.M., Neuhaeuser R.,
  Alcala J.M., Zinnecker H., Wagner R.M., Mundt R., Sterzik M.F. 1996.
  New weak-line T Tauri stars in Taurus-Auriga.
  {\refit Astron.\ Astrophys.\/} 312: 439--454.}

\ref{Wiramihardja, S. D., Kogure T., Yoshida S., Ogura K., Nakano M. 1989.
   Survey observations of emission-line stars in the Orion region.
   I. The Kiso area A-0904.
   {\refit Pub.\ Astron.\ Soc.\ of Japan\/} 41: 155--174.}

\ref{Wiramihardja S.D., Kogure T., Yoshida S., Nakano M., Ogura K.,
   Iwata T. 1991.
   Survey observations of emission-line stars in the Orion region.
   III. The Kiso areas A-0975 and A-0976.
   {\refit Pub.\ Astron.\ Soc.\ of Japan\/} 43: 27--73.}
 
\ref{Wiramihardja S.D., Kogure T., Yoshida S., Ogura K., Nakano M. 1993.
   Survey observations of emission-line stars in the Orion region.
   IV. The Kiso areas A-1047 and A-1048.
   {\refit Pub.\ Astron.\ Soc.\ of Japan\/} 45: 643--653.}

\ref{Witt, A.N. and Schild, R.E. 1986. 
   CCD surface photometry of bright reflection nebulae
   {\refit Astrophys.\ J.\ Supp.\/} 62: 839--852.}

\ref{Wolk, S.J. 1996. {\refit Watching the Stars Go 'Round and 'Round},
   PhD thesis, SUNY Stony Brook, (unpublished).}

\ref{Zapatero Osorio, M.R., Rebolo, R., Martin, E.L. 1996. Brown dwarfs in the
   Pleiades cluster: a CCD-based R, I survey.
   {\refit Astron.\ Astrophys.\/}  317, 164--170.}

\ref{Zinnecker H., McCaughrean M.J., Wilking B.A. 1993. 
    The initial stellar population.
    In {\refit Protostars \& Planets {I}{I}{I}\/}, eds.\ E. H. Levy and
    J. I. Lunine (Tucson: Univ.\ of Arizona Press), pp.\ 429--496}

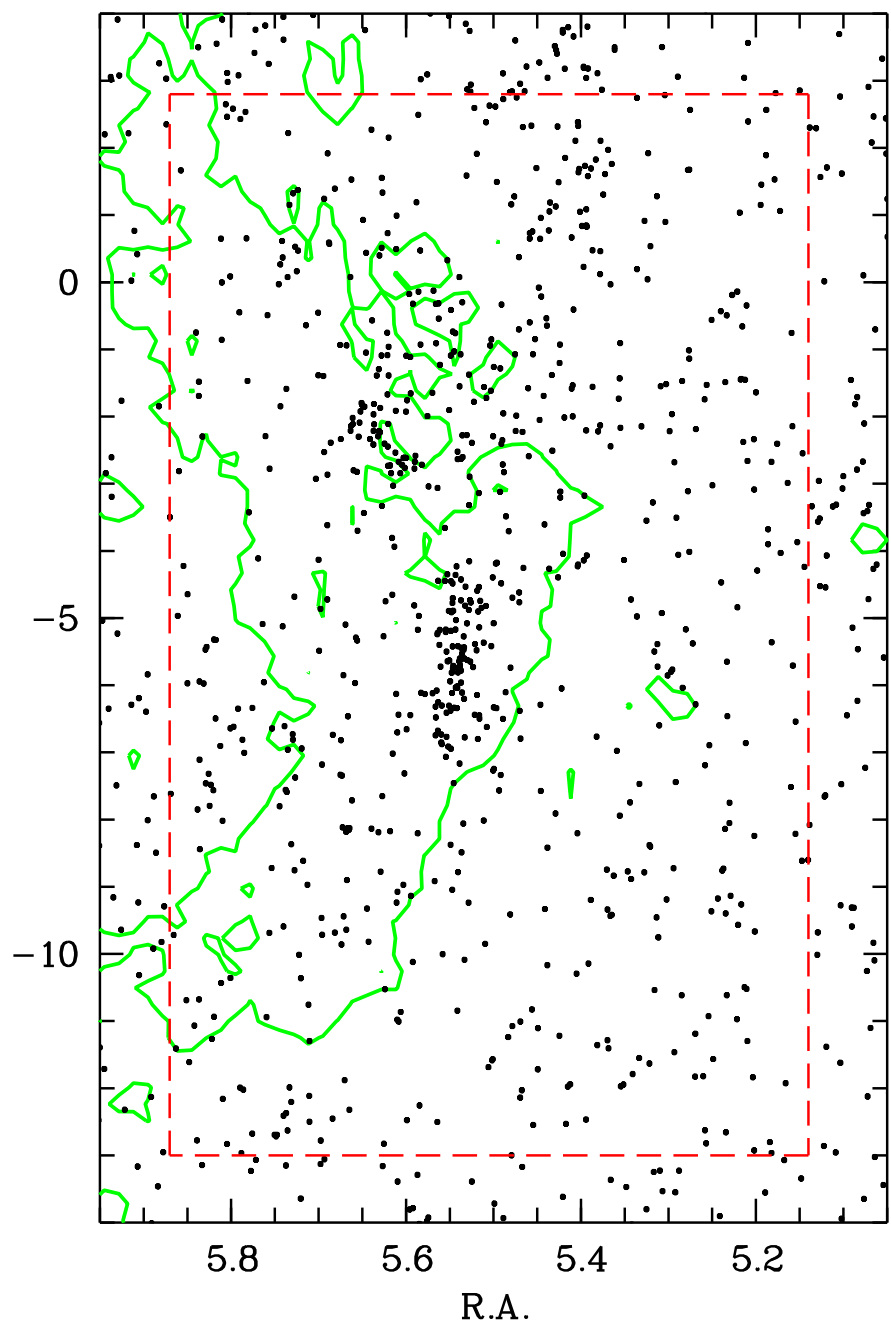
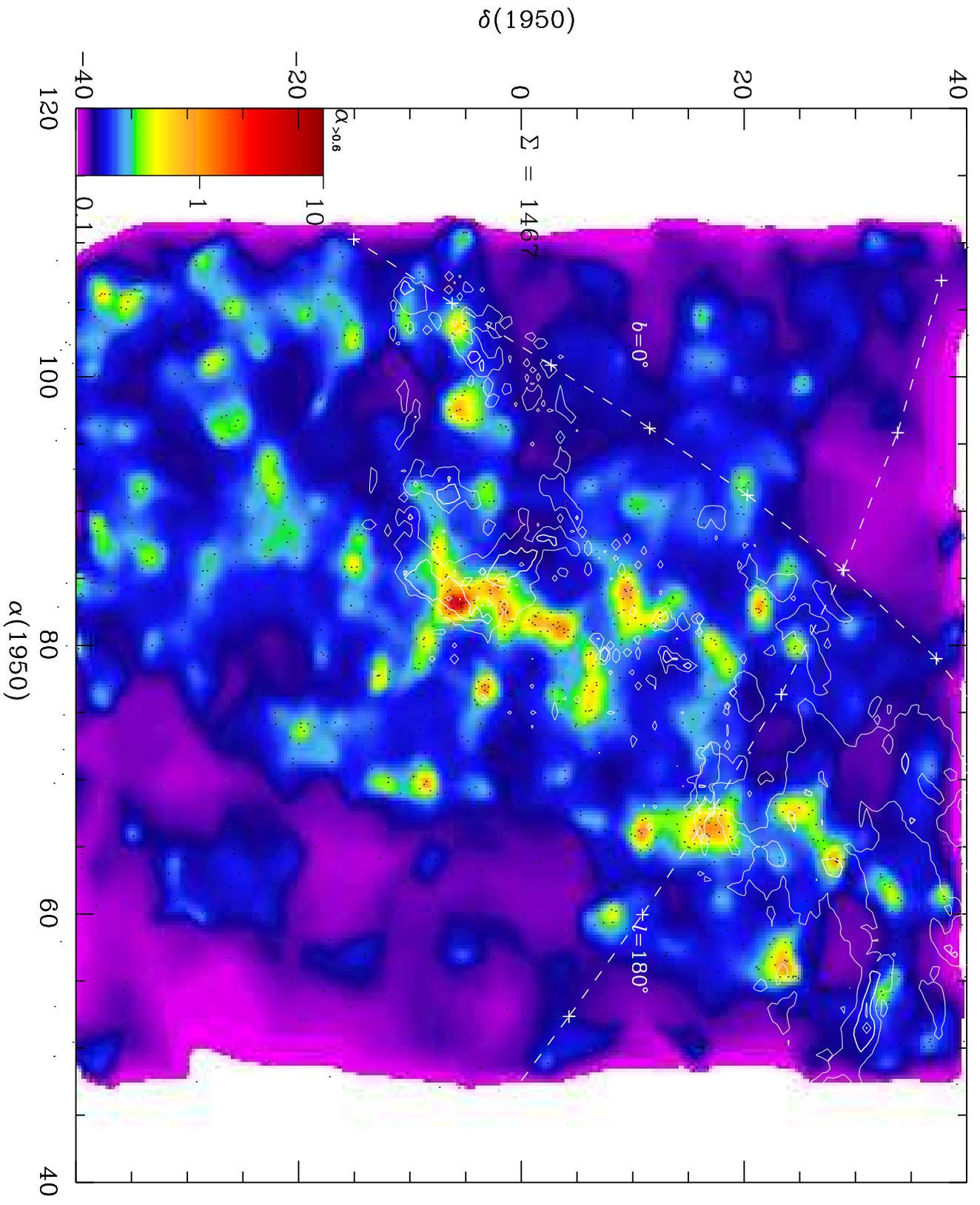
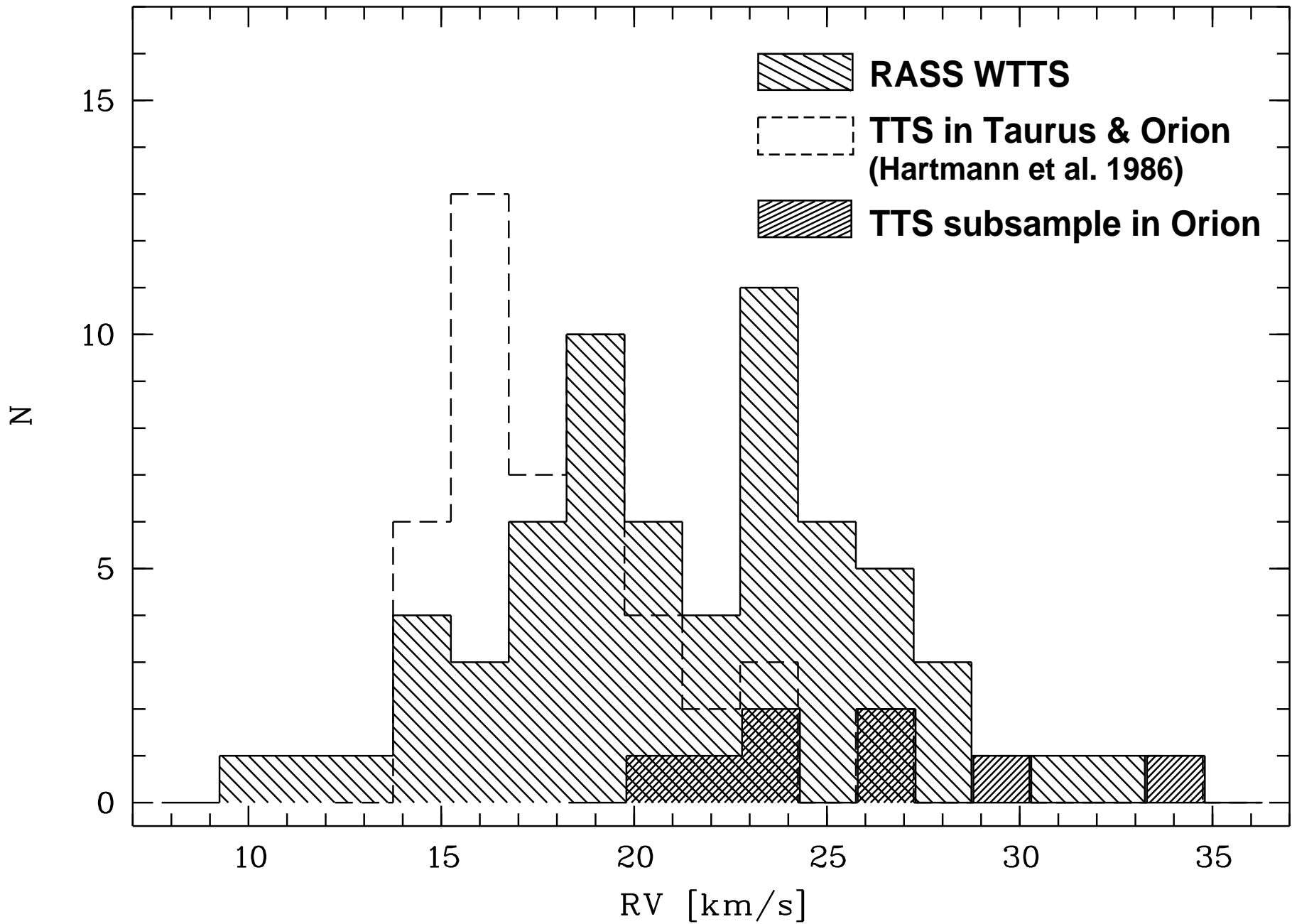
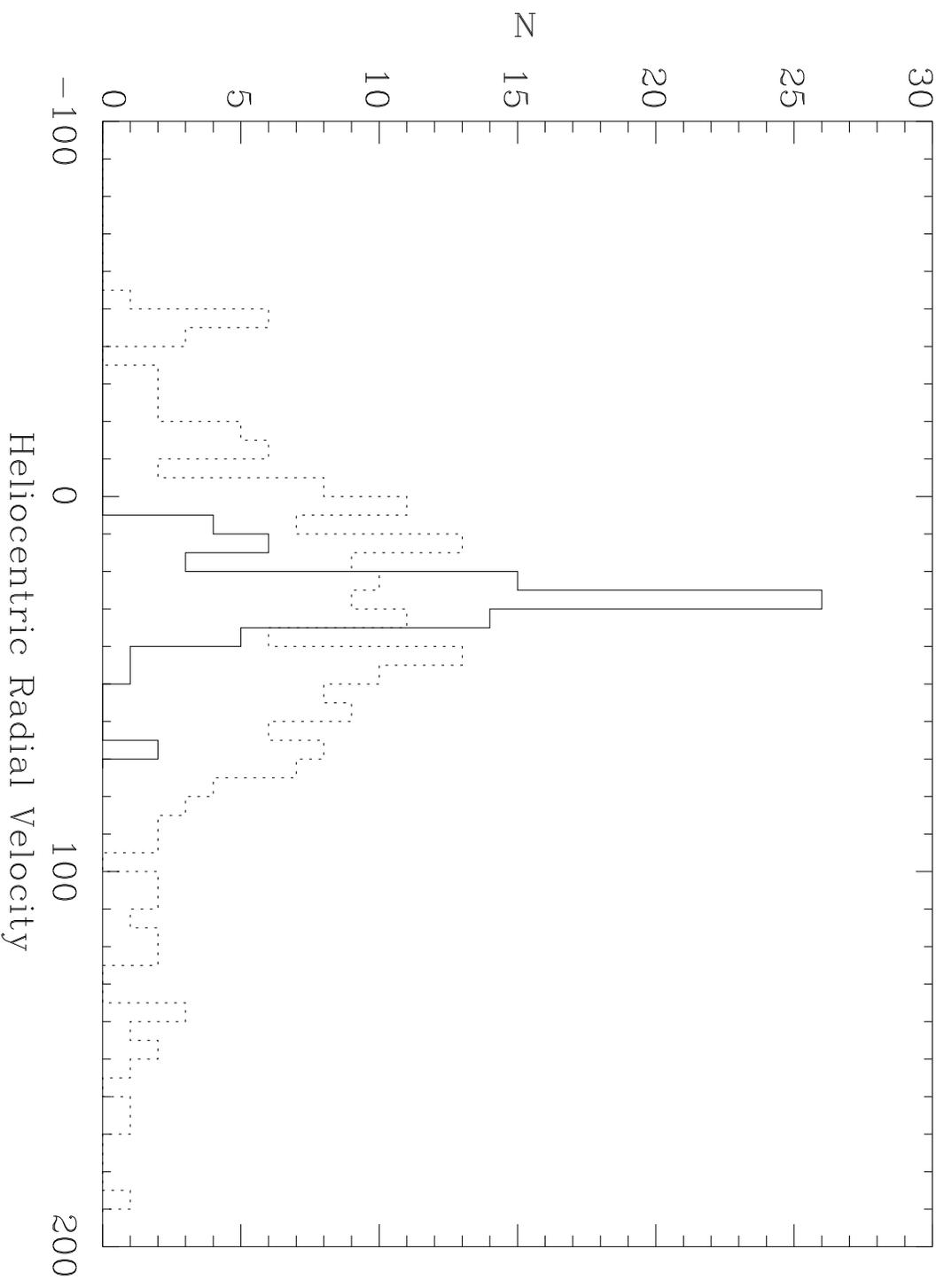
\bye